%% file: main.tex
\documentclass[transmag]{IEEEtran}
\usepackage{latexsym}
\usepackage{graphicx}
\usepackage{amsfonts,amssymb,amsmath}
\usepackage{hyperref}
\usepackage[binary-units]{siunitx}
\usepackage{xcolor}
\DeclareSIUnit{\bps}{bps}
\usepackage{tabularx}
\usepackage{multirow}
\usepackage{booktabs}
\usepackage{diagbox}
\usepackage{flushend}
\usepackage[utf8]{inputenc}
\usepackage[nogroupskip, nopostdot]{glossaries}
\usepackage{glossaries-extra}
\usepackage{fdsymbol}
\makeglossaries
\setabbreviationstyle{long-short}
\loadglsentries{MyAcronym2020}
\setglossarystyle{tree}
\usepackage{color, colortbl}
\definecolor{lred}{rgb}{0.99,0.74,0.71}

\def\BibTeX{{\rm B\kern-.05em{\sc i\kern-.025em b}\kern-.08em T\kern-.1667em\lower.7ex\hbox{E}\kern-.125emX}}
\begin{document}

\title{The Road Towards 6G: A Comprehensive Survey}

\author{Wei~Jiang, \IEEEmembership{Senior Member, IEEE}, Bin~Han, \IEEEmembership{Member, IEEE},\\Mohammad~Asif~Habibi, and Hans~Dieter~Schotten,
\IEEEmembership{Member, IEEE}
\thanks{%Received 21 November 2020; revised 25 January 2021; accepted 31 January 2021.  
\textit{Corresponding author: Wei Jiang  (e-mail: wei.jiang@dfki.de).}}
\thanks{W. Jiang and H. D. Schotten are with the Intelligent Networking Research Group, German Research Center for Artificial Intelligence (DFKI), 67663 Kaiserslautern, Germany (e-mails: \{wei.jiang, hans\_dieter.schotten\}@ dfki.de).}
\thanks{B. Han, M. A. Habibi, and H. D. Schotten are with the Division of Wireless Communications and Radio Navigation (WICON), Department of Electrical and Computer Engineering, University of Kaiserslautern, 67663 Kaiserslautern, Germany (e-mails: \{binhan, asif, schotten\}@eit.uni-kl.de).}}

\IEEEtitleabstractindextext{\begin{abstract}
As of today, the fifth generation (5G) mobile communication system has been rolled out in many countries and the number of 5G subscribers already reaches a very large scale. It is time for academia and industry to shift their attention towards the next generation. At this crossroad, an overview of the current state of the art and a vision of future communications are definitely of interest. This article thus aims to provide a comprehensive survey to draw a picture of the sixth generation (6G) system in terms of drivers, use cases, usage scenarios, requirements, key performance indicators (KPIs), architecture, and enabling technologies. First, we attempt to answer the question of ``Is there any need for 6G?" by shedding light on its key driving factors, in which we predict the explosive growth of mobile traffic until 2030, and envision potential use cases and usage scenarios. Second, the technical requirements of 6G are discussed and compared with those of 5G with respect to a set of KPIs in a quantitative manner.  Third, the state-of-the-art 6G research efforts and activities from representative institutions and countries are summarized, and a tentative roadmap of definition, specification, standardization, and regulation is projected. Then, we identify a dozen of potential technologies and introduce their principles, advantages, challenges, and open research issues. Finally, the conclusions are drawn to paint a picture of ``What 6G may look like?". This survey is intended to serve as an enlightening guideline to spur interests and further investigations for subsequent research and development of 6G communications systems.
\end{abstract}

}
\maketitle

\printunsrtglossaries

\section{INTRODUCTION}
\IEEEPARstart{T}{he} mobile telecommunication industry stems from the first generation (1G) analog cellular systems represented by Advanced Mobile Phone System in the United States and Nordic Mobile Telephone in Europe, which firstly offered mobile voice-calling service around the year 1980. Since then, a new generation of mobile communications was introduced to market nearly every ten years. The 1G analog systems were replaced by the second generation digital cellular networks in around 1990. Despite of several competing systems, the Global System for Mobile Communications known as GSM \cite{Ref_WJ_vriendt2002mobile} achieved a great commercial success and allowed more than one billion of the world's population to enjoy the convenience brought by mobile voice, short texting, and low-rate data services. Exploiting a revolutionary technology named Code-Division Multiple Access (CDMA), the third generation (3G) systems \cite{Ref_WJ_dahlman1998umts} represented by WCDMA, CDMA2000, and TD-SCDMA, were developed and firstly deployed in 2001 to support high-speed data access with a rate of several megabits per second. In December 2009, the commercial Long Term Evolution (LTE) networks \cite{Ref_WJ_astely2009lte} were launched in the Scandinavian capitals Stockholm and Oslo, providing the world's first fourth generation (4G) mobile broadband service. The 4G system that is empowered by a genius combination of multi-input multi-output (MIMO) and orthogonal frequency-division multiplexing (OFDM) spurs the proliferation of smart phones, fostering the mobile Internet industry that is worth trillions of dollars a year. 

In April 2019, when South Korea's three mobile operators and U.S. Verizon were arguing with each other about who is the world's first provider of the fifth generation (5G) communication services, we stepped into the era of 5G. In the past two years, the term of 5G has been remaining one of the hottest buzzwords in the media, attracting unprecedented attention from the whole society. It even went beyond the sphere of technology and economy, becoming the focal point of geopolitical tension. Unlike the previous generations that focused merely on improving network capacities, 5G expands mobile communication services from human to things, and also from consumers to vertical industries. The potential scale of mobile subscription is substantially enlarged from merely billions of the world's population to almost countless inter-connectivity among humans, machines, and things. It enables a wide variety of services from traditional mobile broadband to Industry 4.0, virtual reality (VR), Internet of Things (IoT), and automatic driving \cite{Ref_WJ_andrews2018what}. In 2020, the outbreak of the COVID-19 pandemic leads to a dramatic loss of human life worldwide and imposes unprecedented challenges on societal and economic activities. But this public health crisis highlights the unique role of networks and digital infrastructure in keeping society running and families connected, especially the values of 5G services and applications, such as remote surgeon, online education, remote working, driver-less vehicles, unmanned delivery, robots, smart healthcare, and autonomous manufacturing.  

Currently, 5G is still on its way being deployed across the world, but it is already the time for academia and industry to shift their attention to beyond 5G or the sixth generation (6G) systems, in order to satisfy the future demands for information and communications technology (ICT) in 2030. Even though discussions are ongoing within the wireless community as to whether there is any need for 6G or whether counting the generations should be stopped at 5, adopting the Microsoft’s approach where Windows 10 is the ultimate version, and even there is an opposition to talking about 6G \cite{Ref_WJ_fitzek2020why}, several pioneering works on the next-generation wireless networks have been initiated.  A focus group called \textit{Technologies for Network 2030} within the International Telecommunication Union Telecommunication (ITU-T) standardization sector was established in July 2018. The group intends to study the capabilities of networks for 2030 and beyond \cite{Ref_WJ_li2019blueprint}, when it is expected to support novel forward-looking scenarios, such as holographic-type communications, ubiquitous intelligence, Tactile Internet, multi-sense experience, and digital twin. The European Commission initiated to sponsor beyond 5G research activities, as its recent Horizon 2020 calls - ICT-20 \textit{5G Long Term Evolution} and ICT-52 \textit{Smart Connectivity beyond 5G} – where a batch of pioneer research projects for key 6G technologies were kicked off at the early beginning of 2020. The European Commission has also announced its strategy to accelerate investments in Europe’s ``Gigabit Connectivity" including 5G and 6G to shape Europe's digital future \cite{Ref_WJ_EU2020shaping}. In October 2020, the Next Generation Mobile Networks (NGMN) has launched its new “6G Vision and Drivers” project, intending to provide early and timely direction for global 6G activities. At its meeting in February 2020, the International Telecommunication Union Radiocommunication sector (ITU-R) decided to start study on future technology trends for the future evolution of International Mobile Telecommunications (IMT) \cite{Ref_WJ_itu2020future}.  In Finland, the University of Oulu began ground-breaking 6G research as part of Academy of Finland’s flagship program \cite{Ref_WJ_Aazhang2019key} called 6G-Enabled Wireless Smart Society and Ecosystem (6Genesis), which focuses on several challenging research areas including reliable near-instant unlimited wireless connectivity, distributed computing and intelligence, as well as materials and antennas to be utilized in future for circuits and devices. Besides, other traditional main players in mobile communications, such as the United States, China, Germany, Japan, and South Korea, already initiated their 6G research officially or at least announced their ambitions and tentative roadmaps. At this crossroad, an overview of the current state of the art and a vision of future communications to provide an enlightening guideline for subsequent research and development works is of interest. Recently, the articles focusing on 6G topics, e.g., use cases, application scenarios, requirements, and promising technological pillars, are emerging in the literature, as summarized in the following subsection.

\begin{table*}[!t]
\renewcommand{\arraystretch}{1.3}
\caption{Summary of state-of-the-art contributions related to 6G communication systems}
\label{table_related_works}
\centering
\scriptsize
%\resizebox{\textwidth}{!}{
\begin{tabular}{|c|c|m{1.6cm}|m{12.5cm}|}
\hline
\textbf{Ref.} & \textbf{ Public. Time} & \textbf{Topics} & \textbf{ Major Contributions}    \\  \hline \hline
\cite{Ref_WJ_david20186g}  & Sept. 2018 & Vision & Reviews the key services and innovations from 1G to 5G, and provide a vision for 6G.\\ \hline
\cite{Ref_WJ_nawaz2019quantum} & April 2019 & ML & Reviews the state-of-the-art advances in ML and quantum computing, and proposes a quantum computing-assisted ML framework for 6G networks. \\ \hline 
\cite{Ref_WJ_pappaport2019wireless} & June 2019& THz & Describes the technical challenges and potentials for wireless communications and sensing  above $100\mathrm{GHz}$, and present discoveries, approaches, and recent results that will aid in the development and implementation of 6G networks.   \\ \hline
\cite{Ref_WJ_yang2019wireless}& July 2019 & Vision & Outlines a number of key technological challenges and the potential solutions associated with 6G. \\ \hline
\cite{Ref_WJ_Letaief2019roadmap} & Aug. 2019 & AI &  Discusses potential technologies for 6G to enable ubiquitous AI applications and AI-enabled approaches for the design and optimization of 6G.  \\ \hline
\cite{Ref_WJ_Zong20196g} &Sept. 2019 & Vision & Proposes two candidate system architectures for 6G and identify several 6G technologies including photonics-defined radio, holography, and AI.   \\ \hline 
\cite{Ref_BH_zhang20196g} & Sept. 2019& Survey & A survey aiming to identify requirements, network architecture, key technologies, and new applications.  \\ \hline
\cite{Ref_WJ_Strinati20196g} & Sept. 2019 & Technologies & Five technology enablers for 6G, including pervasive AI at network edge, 3D coverage consisting of terrestrial networks, aerial platforms, and satellite constellation, a new physical layer incorporating sub-\si{\tera\hertz} and VLC, distributed security mechanisms, and a new architecture. \\ \hline
\cite{Ref_WJ_huang2019survey} & Dec. 2019 & Green 6G & A survey on new architectural changes associated with 6G networks and potential technologies such as ubiquitous 3D coverage, pervasive AI, \si{\tera\hertz}, VLC, and blockchain. \\ \hline
\cite{Ref_WJ_jiang2019computational} & Dec. 2019 & AI & A special issue provides a comprehensive treatment on all the technology aspects related to ML for wireless communications, covering fading channel, channel coding, physical-layer design, network slicing, resource management, mobile edge, fog computing, and autonomous network management. \\ \hline
\cite{Ref_WJ_dang2020what} & Jan. 2020 & Vision & Argues that 6G should be human-centric, and therefore security, secrecy, and privacy are key features. To support this vision, a systematic framework, required technologies, and challenges are outlined. \\ \hline
\cite{Ref_WJ_tang2020future} & Feb. 2020 & Vehicular& Summarizes various ML technologies that are promising for communication, networking, and security aspects of vehicular networks, and envisions the ways towards an intelligent 6G vehicular network, including intelligent radio, network intelligentization, and self-learning. \\ \hline
\cite{Ref_WJ_Giordani2020toward}  & Mar. 2020  & Use cases & Foresees several possible use cases and key technologies that are considered as the enablers for these 6G use cases.    \\ \hline
\cite{Ref_WJ_Viswanathan2020communications} & Mar. 2020 & Survey &  New themes including new human-machine interface, ubiquitous computing, multi-sensory data fusion, and precise sensing and actuation. Then, major technology transformations such as new spectrum, new architecture, and new security are presented, with the emphasize of AI's potentials. \\ \hline
\cite{Ref_WJ_zhang2020connecting} & Mar. 2020& AI& Argues that 1~000 times price reduction from the customer’s view point is the key to success and uses AI-assisted intelligent communications to illustrate the drive-force behind.\\ \hline
\cite{Ref_WJ_chen2020vision} & April 2020 & Survey &  A comprehensive discussion of 6G based on the review of 5G developments, covering visions, requirements, technology trends, and challenges, aiming at clarifying the ways to tackle the challenges of coverage, capacity, data rate, and mobility of 6G communication systems.  \\ \hline
\cite{Ref_WJ_saad2020vision} & May 2020 & Survey &  A vision on 6G in terms of applications, technological trends, service classes, and requirements, as well as an identification on enabling technologies and open research problems.   \\ \hline
\cite{Ref_WJ_kato2020challenges} & June 2020& ML& Possible challenges and potential research directions of advancing ML technologies into the future 6G network in terms of communication, networking, and computing perspective. \\ \hline
\cite{Ref_WJ_guo2020explainable} & June 2020 & AI & The core concepts of explainable AI for 6G, including public and legal motivations, definitions, the trade-off between explainability and performance, explainable methods, and an explainable AI framework for future wireless systems. \\ \hline
\cite{Ref_WJ_chowdhury20206g} &Aug. 2020& Survey & A survey on 6G in terms of applications, requirements, challenges, and research directions. Some key technologies such as AI, \si{\tera\hertz}, blockchain, and wireless optical communications are briefly introduced. \\ \hline 
\cite{Ref_WJ_tariq2020speculative} & Aug. 2020 & Vision & Extends the vision of 5G to more ambitious scenarios in a more distant future and speculates on the visionary technologies that can provide the step changes needed for enabling 6G. \\ \hline
\cite{Ref_WJ_liu2020vision} &  Sept. 2020 &  Vision  &  Identifies vision, new application scenarios, and key performance requirements, and proposes a logical mobile network architecture. \\ \hline
\cite{Ref_WJ_gui2020opening} &Oct. 2020& Survey &  A brief presentation of various 6G issues including core services, use cases, requirements, enabling technologies, architectures, typical use scenarios, challenges, and research directions. \\ \hline 
\cite{Ref_WJ_huang2020holographic}&Oct. 2020& MIMO & An overview of holographic MIMO surface (HMIMOS) communications including the available hardware architectures for re-configuring such surfaces, highlighting the opportunities and key challenges in designing HMIMOS-enabled wireless communications for 6G. \\ \hline 
\cite{Ref_WJ_bariah2020prospective} & Oct. 2020 & Survey & A comprehensive study of 6G visions, requirements, challenges, and open research issues, outlining seven disruptive technologies, i.e., millimeter wave (mmWave) communications, \si{\tera\hertz} communications, optical wireless communications, programmable meta-surfaces, drone-based communications, back-scatter communications, and Tactile Internet. \\ \hline
\cite{Ref_WJ_rikkinen2020thz} &  Nov. 2020 & THz  & Analyzes link budget of THz links with justified estimates of calculus terms, such as the achievable or required noise figure, transmit power, and antenna gain. \\ \hline
\cite{Ref_WJ_polese2020toward} &  Nov. 2020 & THz  & Discusses full protocol stack for the realization of end-to-end terahertz 6G mobile networks, from medium access control, network to transport layer.  \\ \hline
\cite{Ref_WJ_wang2020wireless} &  Dec. 2020 & Channel  & A survey of 6G wireless channel measurements and models for full frequency bands, full application scenarios, and full global coverage.   \\ \hline
\cite{Ref_WJ_chi2020visible} &  Dec. 2020 & VLC  & Presents the prospects and challenges of VLC in 6G, its advances in high-speed transmissions, and recent research interests such as new materials and devices, advanced modulation, and underwater VLC.    \\ \hline
\cite{Ref_WJ_kishk2020aerial} &  Dec. 2020 & UAV  & Discusses the advantages of unmanned aerial vehicles to improve the coverage and capacity of 6G, and proposes a network setup utilizing tethered UAVs.  \\ \hline
\cite{Ref_WJ_li2020trustworthy} &  Dec. 2020 & AI  &  Outlines the concept of trustworthy autonomy for 6G and clarifies how explainable AI  can generate the qualitative and quantitative modalities of trust. \\ \hline
\cite{Ref_WJ_du2020machine} &  Dec. 2020 & AI/ML  &  Summarizes some intelligent approaches of applying AI and ML tools to optimize 6G networks, including THz communications,  energy management, security, mobility management, and resource allocation.  \\ \hline

\end{tabular}
\end{table*}

\subsection{State-of-the-Art Related Works}
The earliest article \cite{Ref_WJ_david20186g} that discusses the topic of 6G was published in September 2018, where David and Berndt tried to address the question of ``Is there any need for beyond 5G?" by reviewing the key services and innovations from the 1G analog system to the virtualized and software-defined 5G infrastructure. Nawaz et al. \cite{Ref_WJ_nawaz2019quantum} checked the state-of-the-art advances in the fields of machine learning (ML) and quantum computing, and then envisaged their synergies with communication networks to be considered in the 6G system. In \cite{Ref_WJ_pappaport2019wireless}, Rappaport et al. described the challenges and potentials of terahertz (THz) communications in the development and implementation of 6G networks. Later, the authors of \cite{Ref_WJ_yang2019wireless} provided a brief description of vision and potential techniques. In \cite{Ref_WJ_Letaief2019roadmap}, Letaief et al. discussed potential technologies to enable ubiquitous Artificial Intelligence (AI) service in 6G networks (i.e., networking for AI) and AI-enabled methodologies for the design and optimization of 6G (i.e., AI for networking). Zong et al. proposed two candidate system architectures for 6G in \cite{Ref_WJ_Zong20196g} and identify several 6G technologies including photonics-defined radio, holography, and AI. In the context, the authors of \cite{Ref_BH_zhang20196g} aimed to shed light on requirements, network architecture, key technologies, and new applications in the upcoming 6G system. In \cite{Ref_WJ_Strinati20196g}, Strinati et al. envisioned five technology enablers for 6G, i.e., pervasive AI at network edge, three-dimensional (3D) coverage consisting of terrestrial networks, aerial platforms, and satellite constellation, a new physical layer incorporating sub-\si{\tera\hertz} and Visible Light Communications (VLC), distributed security mechanisms, and a new architecture. Huang et al. surveyed the new architectural changes associated with green 6G networks in \cite{Ref_WJ_huang2019survey}, which also briefly introduces several potential technologies such as ubiquitous 3D coverage, pervasive AI, \si{\tera\hertz}, VLC, and blockchain.  In \cite{Ref_WJ_jiang2019computational}, Jiang and Luo provided a comprehensive and highly coherent treatment on all the technology aspects related to ML for wireless communications and networks.

Since the beginning of 2020, the number of publications related to 6G grows a bit faster than that of the past two years. In \cite{Ref_WJ_dang2020what}, Dong et al. argued that 6G should be human centric, and therefore security, secrecy, and privacy are key features. To support this vision, a systematic framework, required technologies, and anticipated challenges were outlined. Then, a survey on various ML technologies applied for communication, networking, and security aspects of vehicular networks, and a vision of the ways toward an intelligent 6G vehicular network were provided in \cite{Ref_WJ_tang2020future}. Polese et al. foresaw several possible 6G use cases and present a number of technologies, which are considered by the authors as the enablers for these use cases \cite{Ref_WJ_Giordani2020toward}. Viswanathan and Mogensen attempted to  paint a broad picture of communication needs and technologies in the era of 6G \cite{Ref_WJ_Viswanathan2020communications}, where new themes that are likely to shape 6G requirements are presented. Zhang, Xiang, and Xu argued in their article \cite{Ref_WJ_zhang2020connecting} that 1~000 times price reduction from the customer’s view point is the key to success for the 6G system.
In \cite{Ref_WJ_chen2020vision}, Chen et al. contributed a comprehensive discussion that covers visions, requirements, technology trends, and challenges, aiming at clarifying the ways to tackle the challenges of coverage, capacity, data rate, and mobility of 6G mobile communication systems. The authors of \cite{Ref_WJ_saad2020vision} shared their viewpoints in terms of applications, technological trends, service classes, and requirements, and then give their identification on enabling technologies and open research problems. Kato et al. \cite{Ref_WJ_kato2020challenges} recognized possible challenges and potential research directions of advancing ML technologies into the future 6G network from the perspectives of communication, networking, and computing. Guo outlined the core concepts of explainable AI for 6G in \cite{Ref_WJ_guo2020explainable}, including public and legal motivation, definition, the trade-off between explainability and performance, explainable methods, and an explainable AI framework for future wireless systems. A survey paper \cite{Ref_WJ_chowdhury20206g} provides a comprehensive view of 6G in terms of applications, requirements, challenges, and research directions. Some key technologies such as AI, terahertz, blockchain, three-dimensional networking, and wireless optical communications are briefly introduced. In \cite{Ref_WJ_tariq2020speculative}, the authors aimed to extend the vision of 5G to more ambitious scenarios in a more distant future and speculate on the visionary technologies that can provide the step changes needed for enabling 6G. Liu et al. identified the vision of the society development towards 2030 and derive key performance requirements from new applications and services. Taken into account the convergence of information and communication technologies, a logical mobile network architecture is proposed to resolve the lessons from 5G network design \cite{Ref_WJ_liu2020vision}. Guan et al. gave a brief presentation on various 6G issues in \cite{Ref_WJ_gui2020opening}, including core services, use cases, requirements, enabling technologies, architectures, typical use scenarios, challenges, and research directions. The authors of \cite{Ref_WJ_huang2020holographic} provided an overview of holographic MIMO surface communications as a promising technological enabler for 6G wireless communications. 

Recently, Bariah et al. gave a comprehensive 6G vision in \cite{Ref_WJ_bariah2020prospective}, identifying seven disruptive technologies, associated requirements, challenges, and open research issues.  From the viewpoints of radio frequency (RF) hardware and antenna,  \cite{Ref_WJ_rikkinen2020thz} analyzes link budget of THz communication links with justified estimates of calculus terms, such as the achievable or required noise figure, transmit power, and antenna gain. It also evaluates communication distances for links implemented with different technologies and complexity at \SI{300}{\giga\hertz} looking towards anticipated 6G use scenarios. In \cite{Ref_WJ_polese2020toward}, Polese et al. provided an overview of the issues that need to be overcome to introduce the terahertz spectrum in mobile networks, from the perspectives of medium access control, network, and transport layer, with consideration on the performance of end-to-end (E2E) data flows on terahertz connections. Wang et al.  \cite{Ref_WJ_wang2020wireless} provided a survey of 6G wireless channel measurements and models for full frequency bands covering mmWave, THz, and optical wireless communication (OWC) channels, full coverage such as satellite, maritime, and underwater acoustic communication channels, and full application scenarios, e.g., high-speed train, vehicle-to-vehicle, and industry IoT communication channels. Ref. \cite{Ref_WJ_chi2020visible} presents the potential of integrating VLC in 6G, and discusses its technological advances including new materials and devices, modulation, underwater transmission, and ML-based signal processing. The authors of \cite{Ref_WJ_kishk2020aerial} shed light on the advantages of unmanned aerial vehicle (UAV) to improve the coverage and capacity of 6G, and propose a network setup utilizing tethered UAVs. In \cite{Ref_WJ_li2020trustworthy}, the authors outline the concept of trustworthy autonomy for 6G, clarify how explainable AI  can generate the qualitative and quantitative modalities of trust, and provide associated key performance indicators (KPIs) for measuring trust. In \cite{Ref_WJ_du2020machine}, Du et al. summarized some intelligent approaches of applying AI and ML tools to optimize 6G networks, including THz communications,  energy management, security, mobility management, and resource allocation.
To facilitate a clearer illustration, the aforementioned works with major contributions and categorized topics are listed chronologically in Table \ref{table_related_works}. 

\subsection{Contributions}
It is noticed that most of the aforementioned previous works focus merely on one specific aspect of 6G, such as THz \cite{Ref_WJ_pappaport2019wireless}, AI \cite{Ref_WJ_Letaief2019roadmap},  green networks \cite{Ref_WJ_huang2019survey}, use cases \cite{Ref_WJ_Giordani2020toward}, ML \cite{Ref_WJ_kato2020challenges}, and VLC \cite{Ref_WJ_chi2020visible}. There are  a few surveys attempting to provide a complete view, but a \textit{comprehensive} survey is still missing until now. To fill this gap, this article comprehensively surveys the latest advances of 6G research and provides a broad vision in terms of drivers, requirements, efforts, and enablers.
Upon a thorough state-of-the-art analysis of related works, this article is started from envisioning the driving forces, potential use cases, and usage scenarios so as to address the concern on the necessity of developing 6G. Then, the technical requirements needed to support 6G applications and services are clarified in terms of a set of KPIs, and promising technologies are identified and elaborated. The up-to-date research activities across the world are summarized, and the roadmap for research, specification, standardization, and development toward 2030 is projected. Finally, the conclusions are drawn to paint a picture of ``What 6G may look like?".
 
 Compared with existing 6G articles, the main contributions of this article can be listed as follows:
\begin{enumerate}
    \item A thorough state-of-the-art analysis, which provides the most complete summary of related works with latest advances, is made.
    \item It attempts to answer the question of ``Do we really need 6G?" by shedding light on its key drivers, including the explosive growth of mobile traffic and mobile subscriptions until 2030, and disruptive use cases.  It goes beyond the state of the art by identifying the consensus of previous works on use cases  and proposes novel use cases that have never been reported, i.e., Global  Ubiquitous  Connectability, Enhanced  On-Board  Communications, and Pervasive  Intelligence.
    \item Using a holistic methodology, three novel usage scenarios for 6G are proposed, i.e., ubiquitous mobile boardband (uMBB), ultra-reliable low-latency broadband communication (ULBC), and massive ultra-reliable low-latency  communication (mULC). 
    \item It discusses the technical requirements of 6G in terms of a set of KPIs, which are compared with the KPIs of 5G quantitatively, if applicable. 
    \item It summarizes the ambitions, efforts, and research activities on 6G across the world, while a tentative roadmap of definition, specification, standardization, and regulation is envisaged. To the best knowledge of the authors, that is the first time in the literature to provide such an investigation from this perspective. 
    \item An architecture of 3D coverage integrating non-terrestrial and terrestrial networks is envisioned and illustrated within envisioned 6G deployment scenarios.
    \item Unlike all previous works that simply list technological candidates in a line, we categorize 6G enabling technologies into the following groups: New Spectrum, New Networking, New Air Interface, New Architecture, and New Paradigm. This methodology is the first time being used in 6G publications. 
    \item It gives a complete view of potential 6G technologies, which identifies the largest set of enablers by far and the number of identified enablers is far more than any existing survey. The principle, advantages, challenges, and open issues for each enabler are elaborated. Some of the technologies are introduced in detail for the first time from the perspective of 6G, e.g., large-scale satellite constellation and post-quantum security. It includes: \textit{new spectrum} consisting of mmWave, \si{\tera\hertz} communications, VLC, OWC, and dynamic spectrum management (DSM), \textit{new networking} that covers softwarization and virtualization, radio access network (RAN) slicing, open-RAN (O-RAN), and post-quantum security, \textit{new air interface} including massive MIMO, intelligent reflecting surfaces (IRS), coordinated multi-point (CoMP), cell-free massive MIMO, and new modulation techniques, \textit{new architecture} providing 3D coverage by means of integrating large-scale satellite constellation, high-altitude platform (HAP), and UAV with traditional terrestrial networks, and  \textit{new paradigm} empowered by the convergence of computing-communication resources and the integration of mobile networks, AI, blockchain, and  digital twin.
    \item
    It concludes this article by painting a picture of ``What 6G may look like?". The authors envision that 6G would be a radio-optical system, a connected intelligent platform, an integrated space-aerial-terrestrial network, and a smart compute-connect entity to transform the whole Earth into a huge brain, which fully supports the informationized and intelligentized society in 2030 and beyond.
\end{enumerate}

\subsection{Organization of the Article}
The rest of this article is organized as follows: Section II clarifies the key driving forces for the necessity of developing 6G, including the explosive growth of mobile traffic and mobile subscriptions, disruptive use cases, and advanced usage scenarios. Section III analyzes the technical requirements for the 6G system in terms of a number of KPIs. The ambitions and efforts from the main players in the mobile communication industry are summarized and a development roadmap is estimated in Section IV. Section V provides a complete view of a dozen of key technologies for 6G. Finally, Section VI concludes this article by painting a picture about what 6G is.

\section{Drivers} 

Since the middle of 2019, commercial 5G mobile networks have been rolled out across the world and already reached a very large scale in some countries. For example, the number of deployed 5G base stations in China exceeds $500~000$ at the end of 2020, serving more than $100$ million 5G subscribers. Following the tradition that a new generation appears every one decade, it is time for both academia and industry to initiate the exploration of the successor of 5G. On the road towards 6G, however, the first problem we encounter is that there are many concerns like ``\textit{Do we really need 6G?}" or ``\textit{Is 5G already enough?}". To address such questions, we first need to clarify the key driving forces for 6G.  

The development of a next-generation system is driven by not only the exponential growth of mobile traffic and mobile subscriptions but also new disruptive services and applications on the horizon. In addition, it is also driven by the intrinsic need of mobile communication society to continuously improve network efficiencies namely cost efficiency, energy efficiency, spectrum efficiency, and operational efficiency. With the advent of advanced technologies such as AI, \si{\tera\hertz}, and large-scale satellite constellation, the communication network is able to evolve towards a more powerful and more efficient system to better fulfil the requirements of current services and open the possibility for offering disruptive services that have hitherto never been seen. In this section, we intend to shed light on three drivers: \emph{i}) the explosive growth of mobile traffic, \emph{ii})  disruptive use cases, and \emph{iii}) novel usage scenarios.  Its technological drivers will be discussed in detail in Section~\ref{sec:enablers}.

\subsection{Explosively Growing Mobile Traffic}

We are in an unprecedented era where a large number of smart products, interactive services, and intelligent applications emerge and evolve in a prompt manner, imposing a huge demand on mobile communications. It can be foreseen that the 5G system is hard to accommodate the tremendous volume of mobile traffic in 2030 and beyond. Due to the proliferation of rich-video applications, enhanced screen resolution, machine-to-machine (M2M) communications, mobile cloud services, etc., the global mobile traffic will continuously increase in an explosive manner, up to \SI{5016}{\exa\byte}\footnote{1 exabyte (EB)=1~000~000 terabytes (TB), 1 TB=1000 gigabytes (GB)} per month in the year of 2030 compared with 
\SI{62}{\exa\byte} per month in 2020, according to the estimation by ITU-R \cite{Ref_WJ_non2015traffic} in 2015. A report from Ericsson \cite{Ref_WJ_ericsson2020mobile} reveals that the global mobile traffic has reached \SI{33}{\exa\byte} per month at the end of 2019, which justifies the correctness of ITU-R's estimation. 

\begin{figure}[!bpht]
\centering
\includegraphics[width=0.5\textwidth]{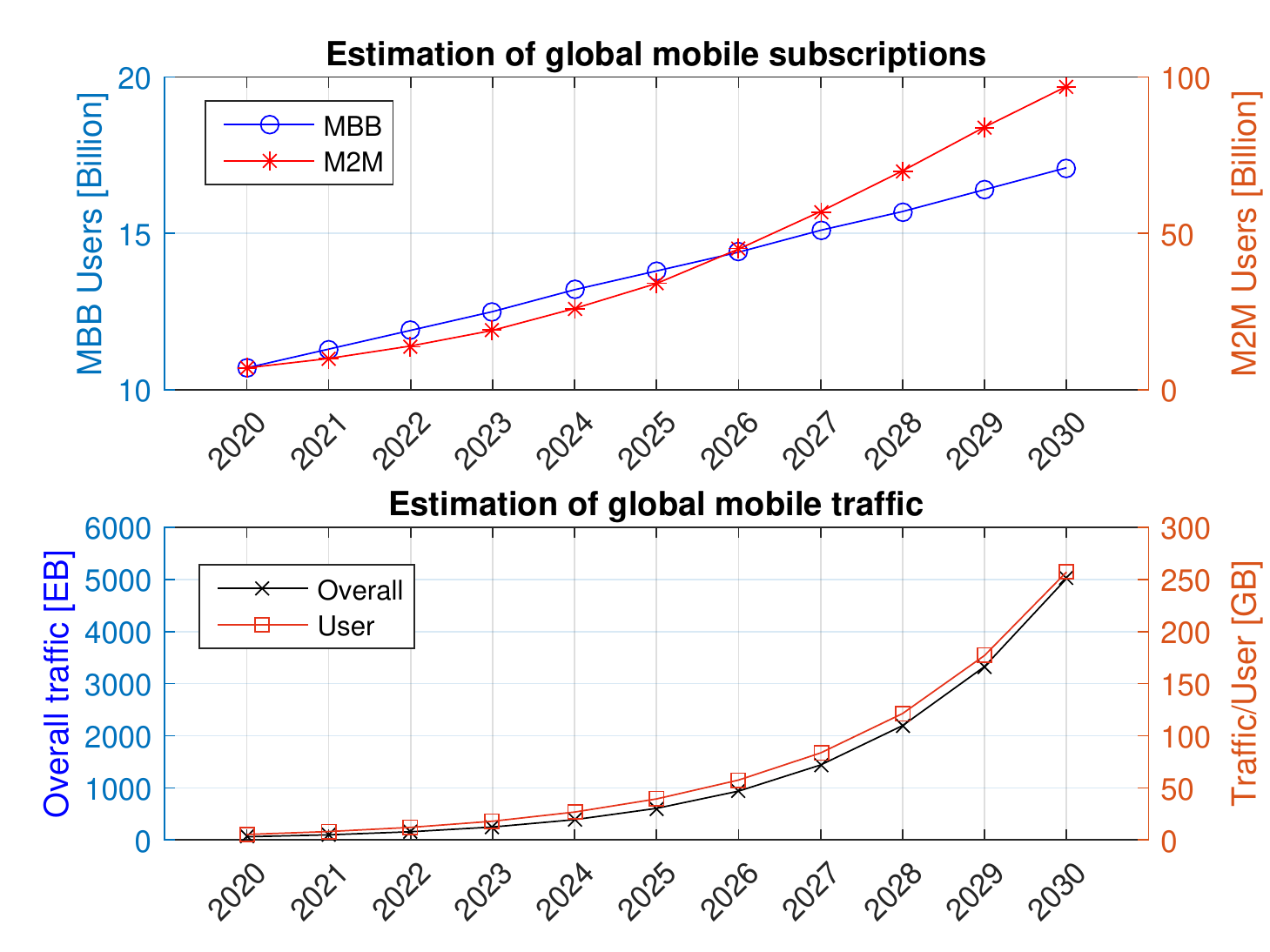}
\caption{Estimated global mobile subscriptions and mobile traffic from 2020 to 2030. Source: ITU-R Report M.2370-0 \cite{Ref_WJ_non2015traffic}. }
\label{Figure_Traffic}
\end{figure}
In the last decade, the number of smartphones and tablets has experienced an exponential growth due to the proliferation of mobile broadband (MBB). This trend will continue in the 2020s since the penetration of smartphones and tablets is not saturated especially in developing countries. Meanwhile,  new-style user terminals, such as wearable electronics and VR glasses, emerge in the market quickly and are adopted by consumers in a fast pace. It is expected that the total number of MBB subscribers worldwide will reach $17.1$ billion by 2030, as shown in \figurename~\ref{Figure_Traffic}. 
On the other hand, the traffic demand per MBB user continuously raises in the company of the rising number of MBB users. That is mainly because of the popularity of mobile video services such as Youtube, Netflix, and more recently Tik-Tok, as well as the stable improvement of screen resolution on mobile devices. The traffic coming from mobile video services already account for two thirds of all mobile traffic nowadays \cite{Ref_WJ_ericsson2020mobile} and is estimated to be more dominant in the future. In some developed countries, a strong traffic growth before 2025 is driven by rich-video services and  a long-term growth wave will continue due to the penetration of augmented reality (AR) and VR applications. The average data consumption for every mobile user per month, as illustrated in \figurename~\ref{Figure_Traffic}, will increase from around \SI{5}{\giga\byte} in 2020 to over \SI{250}{\giga\byte} in 2030. In addition to human-centric communications, the scale of M2M terminals will increase more rapidly and will become saturated no earlier than 2030. It is predicted that the number of M2M subscriptions will reach $97$ billion,  around $14$ times over that of 2020 \cite{Ref_WJ_non2015traffic}. This serves as another driving force for the explosive growth of mobile traffic.

\subsection{Potential Use Cases}
With the advent of new technologies and continuous evolution of existing technologies, e.g., holography, robotics, microelectronics, photo-electronics, AI,  and space technology,  many unprecedented applications can be fostered in mobile networks. To explicitly highlight the unique characteristics and define the technical requirements of 6G, we foresee several representative use cases as follows: 

\textbf{Holographic-Type Communication (HTC)}: Compared to traditional 3D videos using binocular parallax, true holograms can satisfy all visual cues of observing 3D objects by the naked eye as natural as possible.  With a significant advance of holographic display technology in recent years such as  Microsoft's HoloLens \cite{Ref_WJ_hololens}, it is envisioned that its application will become a reality in the next decade. Remote rendering high-definition holograms through a mobile network will bring truly immersive experience. For example, holographic telepresence will allow remote participants to be projected as holograms into a meeting room or allow the attendee of online training or education to interact with ultra-realistic objects. However, HTC  leads to huge bandwidth demands on the order of terabits per second even with image compression. In addition to consider the frame rate, resolution, and color depth in two-dimensional (2D) video, the quality of hologram also involves the volumetric data such as tilt, angle, and position. If representing an object with images every $0.3^{\circ}$, an image-based hologram with $30^{\circ}$ field of view and a tilt of $10^{\circ}$ needs a 2D array of 3300 separate images  \cite{Ref_WJ_clemm2020toward}.  HTC also requires ultra-low latency for true immersiveness and high-precision synchronization across massive bundles of interrelated streams for reconstructing holograms.     

\textbf{Extended Reality (ER)}: Combining augmented, virtual, and mixed realities, ER starts stepping into practical applications in the age of 5G, but it is still in its infancy analogue to the video service at the beginning of mobile Internet. To achieve the same level of image quality,  ER devices with $360^{\circ}$ field of view need much higher data throughput in comparison to 2D video streaming. For an ideal immersion experience, the quality of video with higher resolution, higher frame rate, more color depth, and high dynamic range are required, leading to a bandwidth demand of over \SI{1.6}{\giga\bps} per device \cite{Ref_WJ_huawei2018cloud}. Similar to video traffic that saturates the 4G networks,  the proliferation of ER devices will be blocked by the limited capacity of 5G with the peak rate of \SI{20}{\giga\bps}, especially at the cell edge. In addition to bandwidth, interactive ER applications such as immersive gaming, remote surgery, and remote industrial control, low latency and high reliability are  mandatory. 

\textbf{Tactile Internet}:
It  provides extremely low E2E latency to satisfy the 1-millisecond (\si{\milli\second}) or less reaction time reaching the limit of human sense \cite{Ref_WJ_fettweis2014tactile}. In combination with high reliability, high availability, high security, and sometimes high throughput, a wide range of disruptive real-time applications are enabled. It will play a critical role in the field of real-time monitoring and remote industrial management for Industry 4.0 and Smart Grid. For example, with immersive audio-visual feeds provided by ER or HTC streaming, together with haptic sensing data, a human operator can remotely control the machinery in a place surrounded by biological or chemical hazards, as well as remote robotic surgery carried out by  doctors from hundreds of miles away \cite{Ref_WJ_fettweis2014tactile2}. The typical closed-loop controlling, especially for devices or machinery moving rapidly, is very time-sensitive, where an E2E latency below \SI{1}{\milli\second} is expected.  

\textbf{Multi-Sense Experience}: Human has five senses (sight, hearing, touch, smell, and taste) to perceive external environment, whereas current communications focus only on optical (text, image, and video) and acoustic (audio, voice, and music) media. The involvement of the senses of taste and smell can create fully-immersive experience, which may bring some new services for example in food and texture industries \cite{Ref_WJ_li2019blueprint}. Furthermore, the application of haptic communication will play a more important role and raise a wide range of applications such as remote surgery, remote controlling, and immersive gaming. This use case brings a stringent requirement on low latency. 

\textbf{Digital Twin}
is used to create a comprehensive and detailed virtual copy of a physical (a.k.a. real) object. The softwarized copy is equipped with a wide range of characteristics, information, and properties related to the original object. Such a twin is then used to manufacture multiple copies of an object with full automation and intelligence. The early rollouts of digital twin have attracted significant attention of a number vertical industries and manufacturers. However, its full deployment is expected to be realized with the development of 6G networks.

\textbf{Pervasive Intelligence}: With the proliferation of mobile smart devices and the emergence of new-style connected equipment such as robots, smart cars, drones, and VR glasses, over-the-air intelligent services are envisioned to boom. These intelligent tasks mainly rely on traditional computation-intensive AI technologies: computer vision, simultaneous localization and mapping (SLAM), face and speech recognition, natural language processing, motion control, to name a few. To overcome stringent computation, storage, power, and privacy constrains on mobile devices, 6G networks will offer pervasive intelligence in an AI-as-a-Service manner \cite{Ref_WJ_Letaief2019roadmap} by utilizing distributed computing resources across the cloud, mobile edge, and end-devices, and cultivating communication-efficient ML training and interference mechanisms. For example, a humanoid robot such as Atlas from Boston Dynamics \cite{Ref_WJ_atlas} is possible to off-load its computational load for SLAM towards edge computing resources, in order to improve motion accuracy, prolong battery life, and become more lightweight by removing some embedded computing components. In addition to computation-intensive tasks, pervasive intelligence also facilitates time-sensitive AI tasks to avoid the latency constraint of cloud computing when fast decisions or responses to conditions are required.

\textbf{Intelligent Transport and Logistics}: In 2030 and beyond, millions of autonomous vehicles and drones provide a safe, efficient, and green movement of people and goods. Connected autonomous vehicles have stringent requirements on reliability and latency to guarantee the safety of passengers and pedestrians. Unmanned aerial vehicles, especially swarm of drones, open the possibility for a wide variety of unprecedented applications, while bringing also disruptive requirements for mobile networks.  

\textbf{Enhanced On-Board Communications}:
With the development of economy, the activity sphere of human and the frequency of their movement will rapidly increase in the next decade. The number of passengers travelling by commercial planes, helicopters, high-speed trains, cruise ships, and other vehicles, will be very huge, bringing  skyrocketing demands on high-quality communication services on board. Despite the efforts in the previous generations until 5G, it is undeniable that on-board connectivity is far from satisfactory in most cases due to high mobility, frequent handover, sparse coverage of terrestrial networks, and limited bandwidth and high cost of satellite communications. Relying on reusable space launching technologies and massive production of satellites, the deployment of large-scale satellite constellation such as SpaceX's Starlink \cite{Ref_WJ_starlink}  becomes a reality, enabling cost-efficient and high-throughput  global coverage. Keep this in mind, 6G is expected to be an integrated system of terrestrial networks, satellite constellation, and other aerial platforms to provide seamless 3D coverage, which offers high-quality, low-cost, and global-roaming on-broad communication services.

\textbf{Global Ubiquitous  Connectability}:
The previous generations of mobile communications focused mainly on the dense metropolitan areas, especially indoor scenarios. However, a large population in remote, sparse, and rural areas even have no access to basic ICT services, digging a big digital divide among humans around the world. Besides, more than \SI{70}{\percent} of the Earth's surface is covered by water, where the growth of maritime applications require network coverage for both water surface and underwater. However, ubiquitous coverage across the whole planet with sufficient capacity, acceptable quality of service (QoS), and affordable cost is far from a reality. On the one hand, it is technically impossible for terrestrial networks to cover remote areas and extreme topographies such as ocean, desert, and high mountain areas, while it is too costly to offer terrestrial communication services for sparsely-populated areas. On the other hand, geostationary Earth orbit (GEO) satellites are expensive to deploy and their capacity is limited to several \si{Gbps} per satellite \cite{Ref_WJ_qu2017leo}, which is dedicated only for  high-end users such as maritime and aeronautic industries. As mentioned above, the deployment of large-scale low Earth orbit (LEO) satellite constellation will enable low-cost and high-throughput global communication services \cite{Ref_WJ_hu2001satellite}. The 6G system is envisioned to make use of the synergy of terrestrial networks, satellite constellation, and other aerial platforms to realize ubiquitous connectability for global MBB users and wide-area IoT applications.    

\begin{figure}[!bpht]
\centering
\includegraphics[width=0.45\textwidth]{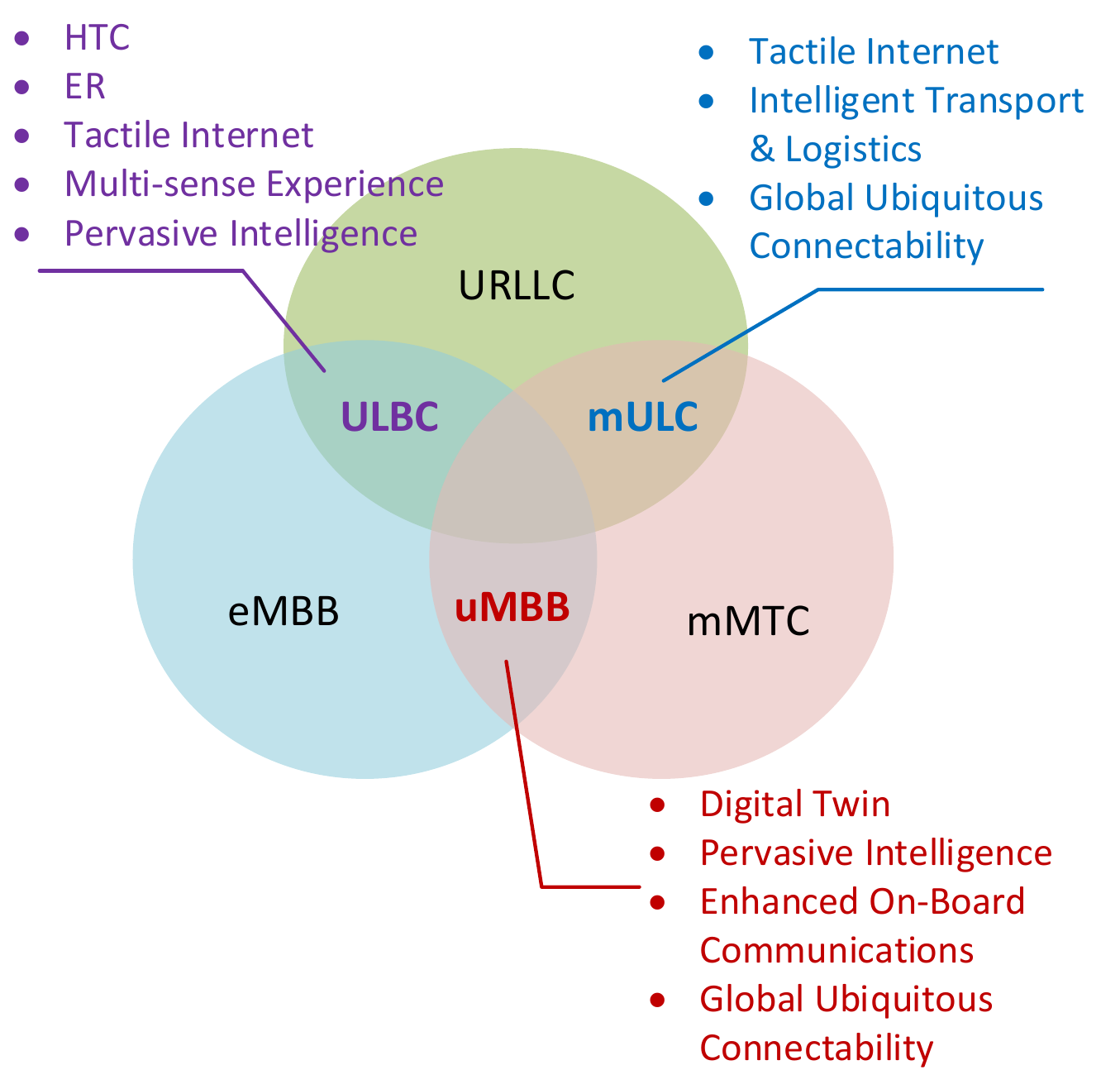}
\caption{In addition to typical 5G usage scenarios (eMBB, ULRRC, and mMTC), three enhanced scenarios named uMBB, ULBC, and mULC are proposed by the authors of this article for the 6G system in order to support disruptive use cases and applications.}
\label{Figure_6GScenarios}
\end{figure}

\begin{figure*}[!t]
\centering
\includegraphics[width=0.88\textwidth]{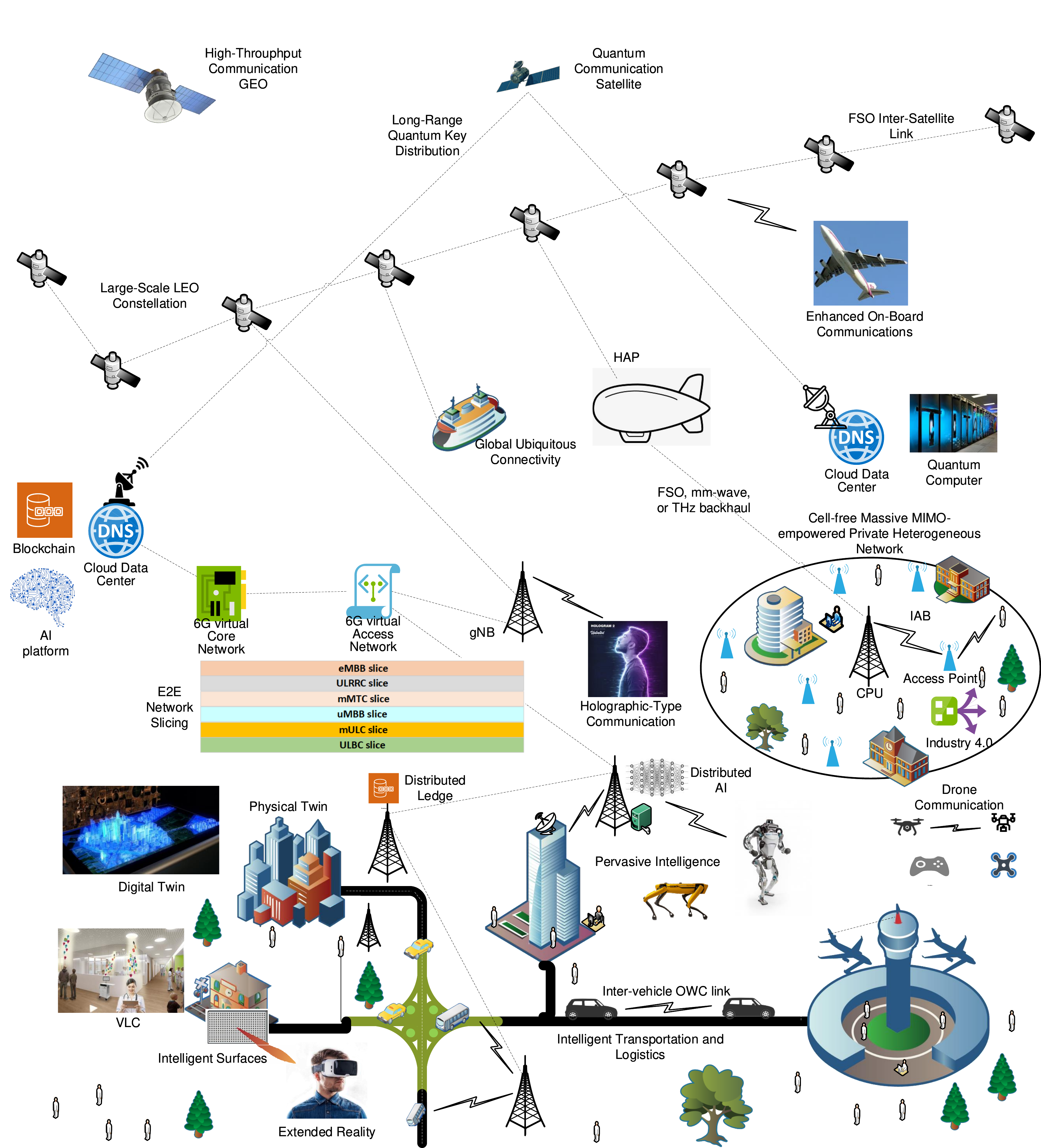}
\caption{An envision of the deployment scenarios and architecture for 6G, demonstrating representative use cases and some of key technological enablers.} 
\label{Figure_6Garchitecture}
\end{figure*}

The connection of the aforementioned use cases with usage scenarios proposed in the next subsection is depicted in \figurename \ref{Figure_6GScenarios}, while some representative use cases are demonstrated within envisaged 6G deployment scenarios, as shown in \figurename \ref{Figure_6Garchitecture}.

\subsection{Usage Scenarios}

The 5G system is designed to meet more diverse QoS requirements arising from a wide variety of vertical applications and services, which have never encountered by mobile subscribers in the previous generations.
To define 5G, three usage scenarios were firstly recommended by ITU-R M.2083 in 2015 \cite{Ref_WJ_non2015imt}:
\begin{itemize}
\item 
enhanced mobile broad-band (eMBB) addresses the human-centric applications for a high-data-rate access to mobile services, multi-media content, and data. This scenario fosters new services and applications over smart devices (smartphones, tablets, and wearable electronics).  It emphasizes wide-area coverage to provide seamless access and high capacity in hot spots.  
\item ultra-reliable low-latency communications (URLLC) is a disruptive promotion over the previous generation systems that focus on human users. It opens the possibility for mission-critical connectivity for new applications such as automatic vehicles, Smart Grid, and Industry 4.0, which have stringent requirements on reliability, latency, and availability.  
\item massive machine-type communications (mMTC) supports dense connectivity with a very large number of connected devices typically deployed in IoT scenarios. The devices such as sensors are low-cost, low-power consumption but typically transmitting a low volume of delay-tolerate data. 
\end{itemize}

It can be seen that these 5G usage scenarios cannot satisfy the technical requirements of the aforementioned 6G use cases. For instance, an user that wears a lightweight VR glass to play interactively-immersive games requires not only ultra-high bandwidth but also low latency. Autonomous vehicles on the road or flying drones need ubiquitous connectivity with high throughput, high reliability, and low latency. Although a few related works discussed potential usage scenarios for 6G, their proposed scenarios are simply an enhancement or extension of 5G scenarios. The definition of scenarios are arbitrary and piece-wise, where the interworking among 6G scenarios and their relations with 5G scenarios are not correctly clarified. In this article, we propose a holistic and more reasonable methodology to define 6G scenarios through extending the scope of current usage scenarios, as shown in \figurename~\ref{Figure_6GScenarios}. Three new scenarios are proposed to meet the requirement of aforementioned use cases, which cover the overlapping areas of 5G scenarios so as to form a complete set. To support high-quality on-board communications and  global ubiquitous connectability, the MBB service should be available across the whole surface of the Earth in the era of 6G, called \textit{ubiquitous MBB} or uMBB. In addition to its ubiquitousness, another enhancement of uMBB is a remarkable boost of network capacity and transmission rate for hot spots so as to support disruptive services, e.g., a group of users wearing lightweight VR glasses gathering in a small room where a data rate of several \si{\giga\bps} per user is needed. The uMBB scenario will be the foundation of digital twin, pervasive intelligence, enhanced on-board communications, and global ubiquitous connectability, as the mapping relationship shown in \figurename~\ref{Figure_6GScenarios}. In addition to KPIs that are applied to evaluate eMBB (such as peak data rate and user-experienced data rate), other KPIs become as same critical as the others in uMBB, i.e., mobility, coverage, and positioning, as indicated in Table \ref{tab:usageScenario_kpis}.    \textit{Ultra-reliable low-latency broadband communication} also called ULBC supports the applications requiring not only URLLC but also extreme high throughput, e.g., HTC-based immersive gaming. It is expected that the use cases of HTC, ER, Tactile Internet, multi-sense experience, and pervasive intelligence will benefit from this scenario. Furthermore, the third scenario called \textit{massive ultra-reliable low-latency  communication} or mULC combines the characteristics of both mMTC and URLLC, which will facilitate the deployment of massive sensors and actuators in vertical industries. Together with eMBB, URLLC, and mMTC, three new scenarios fill the gaps in-between and then a complete set of usage scenarios is formed to support all kinds of use cases and applications in 6G, as shown in \figurename~\ref{Figure_6GScenarios}. Performance requirements (KPIs) required during the design and implementation of such usage scenarios are listed in Table \ref{tab:usageScenario_kpis}.

\section{Requirements}
To well support disruptive use cases and applications in 2030 and beyond, the 6G system will provide extreme capacity, reliability, efficiency, etc. Like the minimal requirements for IMT-2020 specified in \cite{Ref_WJ_non2017minimum}, a number of quantitative or qualitative KPIs are utilized to indicate the technical requirements for 6G. Most of the KPIs that are applied for evaluating 5G are still valid for 6G while some new KPIs would be introduced for the assessment of new technological features. The first eight KPIs in the following part are considered as key requirements in the definition of 5G, which are briefly introduced as follows:

\begin{table*}[!hbtp]
	\centering
	\caption{Performance requirements (KPIs) to support the implementation of usage scenarios. }
	\label{tab:usageScenario_kpis}
	\begin{tabular}{|c|ll|c|c|c|c|c|c|c|c|c|c|c|c|c|c|c|}
		\cline{4-18}
		\multicolumn{3}{l|}{}&\multicolumn{15}{c|}{\textbf{KPI}}\\\hline
		&&&\multirow{11}{*}{\rotatebox[origin=c]{90}{\parbox{33mm}{\raggedright Peak data rate}}}&\multirow{11}{*}{\rotatebox[origin=c]{90}{\parbox{33mm}{\raggedright User-experienced data rate}}}&\multirow{11}{*}{\rotatebox[origin=c]{90}{\parbox{33mm}{\raggedright Latency}}}&\multirow{11}{*}{\rotatebox[origin=c]{90}{\parbox{33mm}{\raggedright Mobility}}}&\multirow{11}{*}{\rotatebox[origin=c]{90}{\parbox{33mm}{\raggedright Connection density}}}&\multirow{11}{*}{\rotatebox[origin=c]{90}{\parbox{33mm}{\raggedright Energy efficiency}}}&\multirow{11}{*}{\rotatebox[origin=c]{90}{\parbox{33mm}{\raggedright Peak spectral efficiency}}}&\multirow{11}{*}{\rotatebox[origin=c]{90}{\parbox{33mm}{\raggedright Area traffic capacity}}}&\multirow{11}{*}{\rotatebox[origin=c]{90}{\parbox{33mm}{\raggedright Reliability}}}&\multirow{11}{*}{\rotatebox[origin=c]{90}{\parbox{33mm}{\raggedright Signal bandwidth}}}&\multirow{11}{*}{\rotatebox[origin=c]{90}{\parbox{33mm}{\raggedright Positioning accuracy}}}&\multirow{11}{*}{\rotatebox[origin=c]{90}{\parbox{33mm}{\raggedright Coverage}}}&\multirow{11}{*}{\rotatebox[origin=c]{90}{\parbox{33mm}{\raggedright Timeliness}}}&\multirow{11}{*}{\rotatebox[origin=c]{90}{\parbox{33mm}{\raggedright Security and privacy}}}&\multirow{11}{*}{\rotatebox[origin=c]{90}{\parbox{33mm}{\raggedright CAPEX and OPEX}}}\\
		&&&&&&&&&&&&&&&&&\\
		&&&&&&&&&&&&&&&&&\\
		&&&&&&&&&&&&&&&&&\\
		&&&&&&&&&&&&&&&&&\\
		&&&&&&&&&&&&&&&&&\\
		\textbf{Generation}&\textbf{Usage Scenario}&&&&&&&&&&&&&&&&\\
		&&&&&&&&&&&&&&&&&\\
		&&&&&&&&&&&&&&&&&\\
		&&&&&&&&&&&&&&&&&\\
		&&&&&&&&&&&&&&&&&\\
		&&&&&&&&&&&&&&&&&\\
		\hline\hline
		\multirow{3}{*}{5G }&\multicolumn{2}{l|}{eMBB}&$\medwhitestar$&$\medwhitestar$&$\checkmark$&$\checkmark$&&$\checkmark$&$\medwhitestar$&$\medwhitestar$&&$\medwhitestar$&&$\checkmark$&&$\checkmark$&$\medwhitestar$\\\cline{2-18}
		&\multicolumn{2}{l|}{URLLC}&&&$\medwhitestar$&&&$\checkmark$&&&$\medwhitestar$&$\checkmark$&$\checkmark$&$\medwhitestar$&$\medwhitestar$&$\medwhitestar$&\\\cline{2-18}
		&\multicolumn{2}{l|}{mMTC}&&&&&$\medwhitestar$&$\medwhitestar$&&&&&&$\checkmark$&&$\checkmark$&\\ \hline
		\multirow{3}{*}{6G }&\multicolumn{2}{l|}{uMBB}&$\medwhitestar$&$\medwhitestar$&$\checkmark$&$\medwhitestar$&&$\checkmark$&$\medwhitestar$&$\medwhitestar$&$\checkmark$&$\medwhitestar$&$\checkmark$&$\medwhitestar$&&$\checkmark$&$\medwhitestar$\\\cline{2-18}
		&\multicolumn{2}{l|}{mULC}&&&$\medwhitestar$&&$\medwhitestar$&$\medwhitestar$&&&$\medwhitestar$&$\checkmark$&$\checkmark$&$\medwhitestar$&$\medwhitestar$&$\medwhitestar$&\\\cline{2-18}
		&\multicolumn{2}{l|}{ULBC}&$\medwhitestar$&$\medwhitestar$&$\medwhitestar$&$\checkmark$&&$\checkmark$&$\medwhitestar$&$\medwhitestar$&$\medwhitestar$&$\medwhitestar$&$\checkmark$&$\medwhitestar$&$\medwhitestar$&$\medwhitestar$&\\ \hline 		\multirow{2}{*}{\textbf{Legend}}&\multicolumn{17}{l|}{$\checkmark$:\quad Generic/weak impact}\\
		&\multicolumn{17}{l|}{$\medwhitestar$:\quad Specialized/critical impact}\\\hline
	\end{tabular}
\end{table*}

\begin{itemize}
\item \textbf{Peak data rate} is the highest  data rate under ideal conditions, in which all available radio resources are totally assigned to a single mobile station. Traditionally, it is the most symbolic parameter to differentiate different generations of mobile systems. Driven by both user demand and technological advances such as \si{\tera\hertz} communications, it is expected to reach up to \SI{1}{\tera\bps}, tens of times that of 5G, which has the peak rate of \SI{20}{\giga\bps} for downlink and \SI{10}{\giga\bps} for uplink. 
\item \textbf{User-experienced data rate} is defined as the $5^{th}$ percentile point ($5\%$) of the cumulative distribution function of user throughput. In other words, a user can get at least this data rate at any time or location with a possibility of $95\%$. It is more meaningful to measure the perceived performance, especially at the cell edge, and reflect the quality of network design such as site density, architecture, inter-cell optimization, etc. In the 5G deployment scenario of dense urban, the target of user-perceived rate  is \SI{100}{\mega\bps} for downlink and \SI{50}{\mega\bps} for uplink. It is expected that 6G can offer \SI{1}{\giga\bps} even higher, which is 10 times that of 5G.
\item \textbf{Latency} can be differentiated into user plane and control plane latency. The former is the time delay induced in a radio network from a packet being sending out at the source until the destination receives it, assuming a mobile station is in the active state. In 5G, the minimum requirement for user plane latency is \SI{4}{\milli\second} for eMBB and \SI{1}{\milli\second} for URLLC. This value is envisioned to be further reduced to  \SI{100}{\micro\second} or even \SI{10}{\micro\second}. Control plane latency refers to the transition time from a most “battery efficient” state (e.g., the idle state) to the start of continuous data transfer (e.g., the active state). The minimum latency for control plane should be \SI{10}{\milli\second}  in 5G and is expected to be also remarkably improved in 6G. In addition to over-the-air delay,  round-trip or E2E delay  is more meaningful but also complicated due to the large number of network entities involved. In 6G, the E2E latency may be considered as a whole.
\item \textbf{Mobility} means the highest moving speed of a mobile station supported by a network with the provisioning of acceptable Quality of Experience (QoE). To support the deployment scenario of high-speed trains, the highest mobility supported by 5G is \SI{500}{\kilo\meter/\hour}.  In 6G, the maximal speed of \SI{1000}{\kilo\meter/\hour} is targeted if commercial airline systems are considered.  
\item \textbf{Connection density}  is the KPI applied for the purpose of evaluation in the  usage scenario of mMTC. Given a limited number of radio resources, the minimal number of devices with a relaxed QoS per square kilometer (\si{\kilo\meter\squared}) is $10^6$ in 5G, which is envisioned to be further improved $10$ times to $10^7$ per \si{\kilo\meter\squared}.
\item \textbf{Energy efficiency} is important to realize cost-efficient mobile networks and reduce the total carbon emission for green ICT, playing a critical role from the societal-economic respective. After the early deployment of 5G networks, there is already some complaints about its high energy consumption although the energy efficiency per bit has been substantially improved in comparison with the previous generations. In 6G networks, this KPI would be $10 - 100$ times better over that of 5G so as to improve the energy efficiency per bit while reducing the overall power consumption of the mobile industry.
\item \textbf{Peak spectral efficiency} is an important KPI to measure the advance of radio transmission technologies. The minimum requirement in 5G for peak spectral efficiencies are \SI{30}{\bps/\hertz} in the downlink and \SI{15}{\bps/\hertz} in the uplink. Following the empirical data, it is expected that advanced 6G radio technologies can achieve three times higher spectral efficiency over the 5G system. 
\item \textbf{Area traffic capacity} is a measurement of the total mobile traffic that a network can accommodate per unit area, relating to the available bandwidth, spectrum efficiency, and network densification. The minimal requirement for 5G is \SI{10}{\mega\bps} per square meter (\si{\meter\squared}), which is expected to reach \SI{1}{\giga\bps/\meter\squared} in some deployment scenarios such as indoor hot spots. 
\end{itemize}

In addition to the aforementioned key capabilities, there are several extended or novel KPIs that may be also required so as to properly evaluate the requirements of 6G. 
\begin{itemize}
\item \textbf{Reliability} relates to the capability of transmitting a given amount of traffic within a predetermined time duration with high success probability. This requirement is defined for the purpose of evaluation in the  usage scenario of URLLC. 
In 5G networks, the minimum requirement for the reliability is measured by a success probability of $1-10^{-5}$ when transmitting a data packet of $32$ bytes within \SI{1}{\milli\second} given the channel quality of coverage edge for the deployment scenario of urban macro environment. It is expected to improve at least two orders of magnitude, i.e., $1-10^{-7}$  or \SI{99.99999}{\percent} in the next-generation system.
\item \textbf{Signal bandwidth} is the maximum aggregated system bandwidth. The bandwidth may be supported by single or multiple RF carriers. The requirement for bandwidth in 5G is at least \SI{100}{\mega\hertz}, and 6G will support up to \SI{1}{\giga\hertz} for operation in higher frequency bands or even higher in \si{\tera\hertz} communications or OWC. 
\item \textbf{Positioning accuracy} of the 5G positioning service is better than \SI{10}{\meter}. Higher accuracy of positioning has a strong demand in many vertical and industrial applications, especially in indoor environment that cannot be covered by satellite-base positioning systems. With the application of \si{\tera\hertz} radio station, which has a strong potential in high-accuracy positioning, the accuracy supported by 6G networks is expected to reach centimeter (\si{\centi\meter}) level. 
\item \textbf{Coverage} in the definition of 5G requirement mainly focuses on the received quality of radio signal within a single base station. The coupling loss, which is defined as the total long-term channel loss over the link between a terminal and a base station and includes antenna gains, path loss, and shadowing, is utilized to measure the area served by a base station. In 6G networks, the connotation of coverage should be substantially extended considering that the coverage will be globally ubiquitous and will be shifted from only 2D in terrestrial networks to 3D in a terrestrial-satellite-aerial integrated system.   
\item \textbf{Timeliness} is an emerging time-domain performance requirement to future communication systems. Typical metrics of timeliness include the well-known age-of-information (AoI)~\cite{Ref_BH_kosta2017age}, and its recently proposed variants such like age-of-task (AoT)~\cite{Ref_BH_song2019age} and age-of-synchronization (AoS)~\cite{Ref_BH_tang2020scheduling}. Differing from the classical memoryless metric of latency, which focuses on the overall delay experienced by all data packets or service sessions throughout their delivery process, the concept of timeliness emphasizes the freshness of the latest data and service that are successfully delivered to the end user. This brings to the system an endogenous birth-time discrimination against outdated data/service, as well as a memory to its historical state(s), and therewith raises both the impact and the complexity of task scheduling in system optimization.
{
\item \textbf{Security and privacy} are necessary  for assessing whether the operation of a network is secure enough to protect infrastructure, devices, data, and assets. The main security tasks for mobile networks are \textit{confidentiality} that prevents sensitive information from being exposing to unauthorized entities, \textit{integrity} guaranteeing that information is not modified illegally,  and \textit{authentication}  ensuring that the communicating parties are who they say they are.  On the other hand, privacy becomes a high priority to address growing concern and privacy legislation such as the General Data Protection Regulation (GDPR) in Europe. Some KPIs can be applied to quantitatively measure security and privacy, e.g., percentage of security threats that are identified by threat identification algorithms, with which the effectiveness of anomaly detection can be evaluated. 
\item \textbf{Capital and operational expenditure}}  is a critical factor to measure the affordability of mobile services, influencing substantially the commercial success of a mobile system. The expenditure of a mobile operator can be divided into two main aspects: capital expenditure (CAPEX) that is the cost spent to build communication infrastructure and operational expenditure (OPEX) used for maintenance and operation. Due to the network densification, mobile operators suffer from a pressure of high CAPEX. Meanwhile, mobile networks' troubleshooting (systems failures, cyber-attacks, and performance degradations, etc.) still cannot avoid manual operations. A mobile operator has to keep an operational group with a large number of network administrators with high expertise, leading to a costly OPEX that is currently three times that of CAPEX and keeps rising \cite{Ref_WJ_jiang2017experimental}. During the design of 6G, the expenditure will be a key factor to consider.
\end{itemize}
\begin{figure}[!tbph]
\centering
\includegraphics[width=0.48\textwidth]{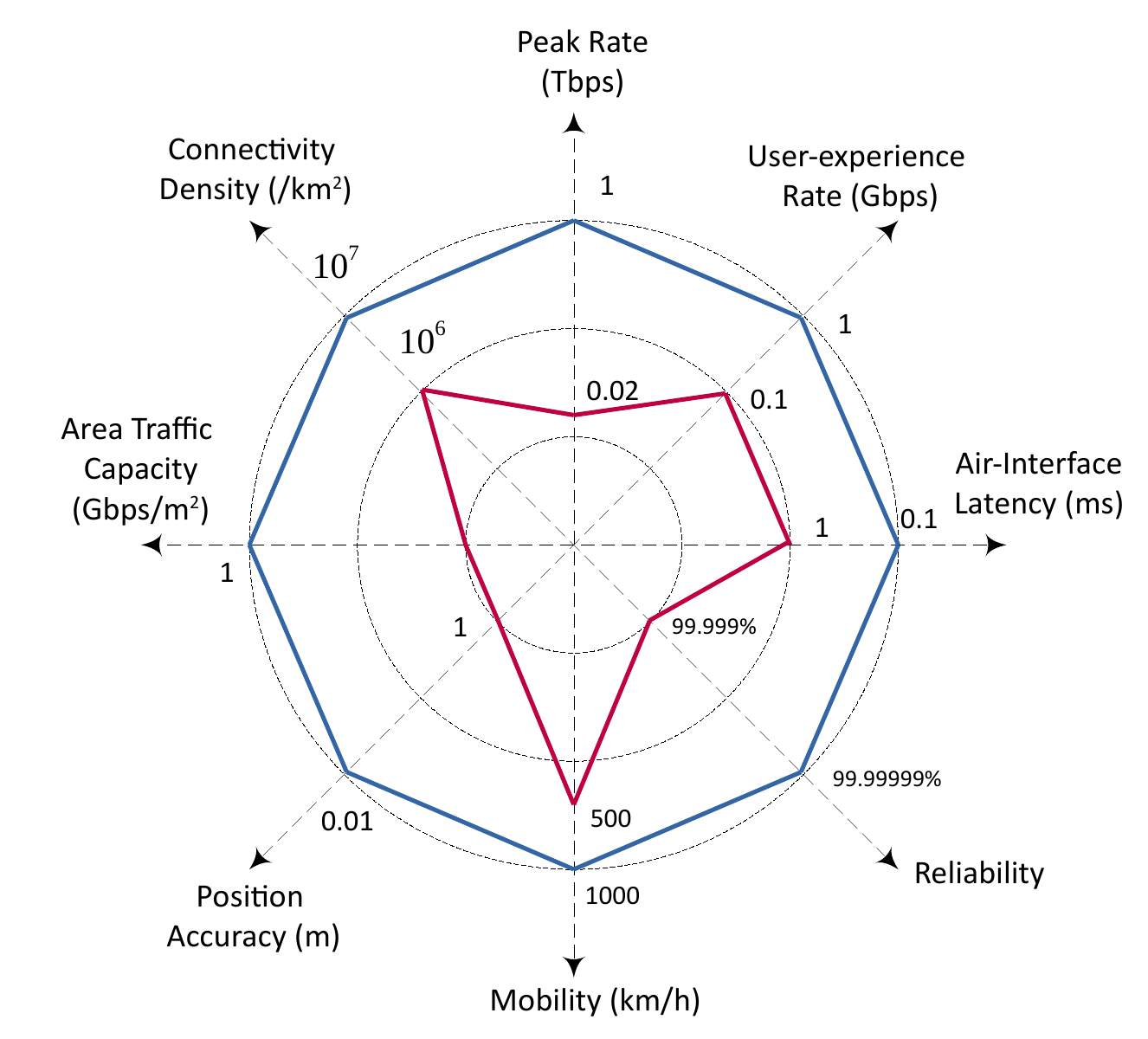}
\caption{Quantitative comparison of the technical requirements between 5G and 6G w.r.t eight representative KPIs. The vertices of the inner polygon stands for the KPIs for 5G, while the vertices of the outer octagon represent that of 6G. Different dotted circles indicate quantities in an exponential manner instead of proportional scale, where the value in a bigger circle stands for one order of magnitude ``better" than that of the neighboring smaller circle. For example, the minimal latency of 5G is defined as \SI{1}{\milli\second} in comparison with  \SI{0.1}{\milli\second} expected in 6G, amounting to 10 times better, while the peak rate of 6G is envisioned to be \SI{1}{\tera\bps} that is 50 times over 5G.   }
\label{Figure_5Gflower}
\end{figure}
To provide a quantitative performance comparison between 5G and 6G, eight representative KPIs are visualized, as shown in \figurename \ref{Figure_5Gflower}. The KPIs required to assess 5G and 6G usage scenarios and the enabling technologies that can support the implementation of each KPI are summarized in Table \ref{tab:usageScenario_kpis} and Table \ref{tab:enablers_kpis}, respectively.

\section{Roadmap and Efforts} 
\begin{figure*}[!t]
\centering
\includegraphics[width=0.99\textwidth]{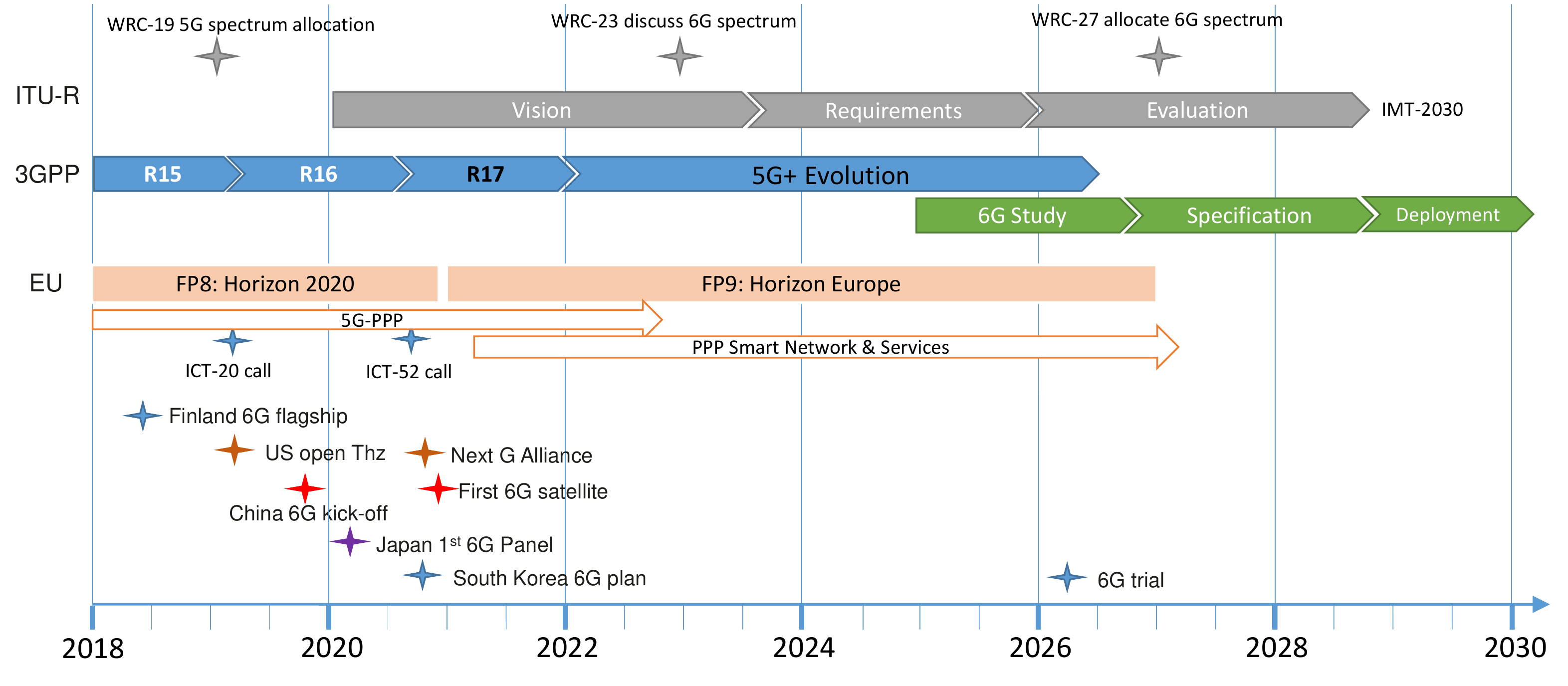}
\caption{The predicted roadmap of research, definition, specification, spectrum regulation, development, and deployment for 6G mobile systems.} 
\label{Figure_6GTimeline}
\end{figure*}
Even though discussions are ongoing within the wireless community as to whether there is any need for 6G and whether counting the generations should be stopped at 5, a few pioneering works on the next-generation wireless networks have been initiated. In this section, the up-to-date advances on 6G research from representative institutions and countries are summarized, while a tentative roadmap of definition, specification, standardization, and regulation is projected, as demonstrated in \figurename  \ref{Figure_6GTimeline}.  In July 2018, a focus group ``\textit{Technologies for Network 2030}" was established under ITU-T \cite{Ref_WJ_li2019blueprint}. This group intends to study and review existing technologies, platforms, and standards for identifying the gaps and challenges towards the capabilities of networks for the year 2030 and beyond, with the emergence of novel forward-looking scenarios such as holographic applications, Tactile Internet, multi-sense networks, and digital twin. Although it mainly focuses on fixed data communication networks,  the vision, requirements, architecture, and novel use cases identified in this group also has reference values for the definition of the 6G mobile system. 
According to the empirical timeline, ITU-R section will initiate the study of 6G vision and will publish the requirements for IMT for 2030 (as the requirements of IMT-2020 \cite{Ref_WJ_non2017minimum} published in 2017) in the middle of the 2020s, and then will step into the evaluation phase afterwards. Considering the successful accomplishments by ITU for the evolution of IMT-2000, IMT-Advanced and IMT-2020, similar actions are proposed for the evolution of IMT towards 2030 and beyond. At its meeting in February 2020, ITU-R working party 5D decided to start study on future technology trends for the future evolution of IMT \cite{Ref_WJ_itu2020future}. It is planned to complete this study at the meeting in June 2022. A preliminary draft report will be developed and will consider related information from various external organizations and country/regional research programs.  
ITU-R is also responsible for organizing the world radiocommunication conference (WRC) that governs the frequency assignment, being hold every three to four years.  In WRC-19 held in 2019, the spectrum allocation issue for the 5G system was determined. It is expected that the WRC probably scheduled in 2023 (WRC-23) will discuss the spectrum issues for 6G and the spectrum allocation for 6G communications may be formally decided in 2017 (WRC-27). 

In the early of 2019, the third generation partnership project (3GPP) has frozen the Release 15 (Rel. 15 or R15) specifications, which are the first phase of 5G standards. Rel. 15 mainly focuses on eMBB and provides the basis for URLLC, especially in respect of the support of low latency.  In July 2020, the subsequent release (i.e., Rel. 16) has been completed as the second phase of 5G standards \cite{Ref_WJ_ghosh2019evolution}. In addition to enhancements for the existing Rel. 15 features, new features such as non-public network, new radio (NR) unlicensed, NR positioning, NR-Light, and integrated access and backhaul (IAB) have been introduced so as to fully support URLLC and industrial IoT. Currently, a more advanced version (Rel. 17) is being standardized by 3GPP and is expected to be completed in the early of 2021, in spite of a delay due to the COVID-19 pandemic. Driven by a multitude of key stakeholders from the traditional mobile industry, a wide range of verticals, and the non-terrestrial industry, it is envisioned as the most versatile release in 3GPP history in terms of features content, including NR over non-terrestrial networks, NR beyond \SI{52.6}{\mega\hertz}, NR sidelink enhancement, network automation, etc. 3GPP is expected to standardize several subsequent releases to further evolve the 5G system, which can be called 5G+ or 5G Evolution. According to the experiences got in previous generations, 6G will be a disruptive system that cannot be developed following such a backward-compatible manner. In parallel, therefore, 3GPP is expected to initiate the study item for 6G around the year 2025, followed by the phase of specification, to guarantee the first commercial deployment roll-out of 6G by 2030.  

In October 2018, the European Commission has initiated to sponsor beyond 5G research activities by opening the ICT-20-2019 call ``\textit{5G Long Term Evolution}" under the eighth Framework Programmes for Research and Technological Development  (FP8) being named \textit{Horizon 2020}. Eight projects such as 5G-COMPLETE \cite{Ref_WJ_5gcomplete} and 5G-CLARITY \cite{Ref_WJ_5gclarity} have been selected from a total of 66 proposals and kicked off in  early 2020. In its recent call ICT-52-2020 ``\textit{Smart Connectivity beyond 5G}", the accepted projects selected from a high-competitive evaluation process explicitly shown that their ambition is to provide the early research efforts on 6G. The details of ICT-20 and ICT-52 research projects are summarized in Table \ref{tab:EU_projects}. In its upcoming research and innovation framework program called Horizon Europe or FP9, a large number of efforts and funding will concentrate on the research and development of 6G and will be organized in the framework of Public Private Partnership (PPP) ``\textit{Smart Network \& Services}", following the successful strategy of the 5G Infrastructure Public Private Partnership (5G-PPP) under Horizon 2020. Furthermore, the European Commission has  announced in February 2020 its strategy to accelerate investments in Europe’s ``Gigabit Connectivity" including 5G and 6G to shape Europe's digital future \cite{Ref_WJ_EU2020shaping}.

\begin{table*}[!t]
\renewcommand{\arraystretch}{1.3}
	\caption{Summary of EU beyond-5G and 6G research projects}
	\label{tab:EU_projects}
	\scriptsize
\begin{tabular}{|c|l|m{9cm}|m{4.8cm}|}
\hline
\textbf{Call} & \textbf{Acronym} & \textbf{Project Title} & \textbf{ Major Research Topics}    \\  \hline 
\multirow{14}{*}{ICT-20}  & 5G-CLARITY & Beyond 5G multi-tenant private networks integrating Cellular, WiFi, and LiFi, Powered by ARtificial Intelligence and Intent Based PolicY & Private networks, AI-driven network automation, intent-based network \\ \cline{2-4}
 &  5G-COMPLETE &A unified network, Computational and stOrage resource Management framework targeting end-to-end Performance optimization for secure 5G muLti-tEchnology and multi-Tenancy Environments &  Computing-storage-network convergence, architecture, post-quantum cryptosystem, fiber-wireless fronthaul\\ \cline{2-4}
 & 5G-ZORRO & Zero-touch security and trust for ubiquitous computing and connectivity in 5G networks &  Security, privacy, distributed ledge technology (DLT), zero-touch automation, E2E network slicing\\ \cline{2-4}
 & ARIADNE & Artificial Intelligence Aided D-band Network for 5G Long Term Evolution & D-band, metasurfaces, AI-based management\\ \cline{2-4}
 & INSPIRE-5G+ & INtelligent Security and PervasIve tRust for 5G and Beyond & Trusted multi-tenancy, security, AI, blockchain\\ \cline{2-4}
 & LOCUS & LOCalization and analytics on-demand embedded in the 5G ecosystem, for Ubiquitous vertical applicationS & Localization and analytics, location-as-a-service\\ \cline{2-4}
 & MonB5G & Distributed management of Network Slices in beyond 5G & Network slice management, zero-touch automation, AI-assisted security \\ \cline{2-4}
 & TERAWAY & Terahertz technology for ultra-broadband and ultra-wideband operation of backhaul and fronthaul links in systems with SDN management of network and radio resources & THz, photonics-defined transceiver, bachhaul and fronthaul, network and resource management\\ \hline
\multirow{16}{*}{ICT-52} & 6G BRAINS & Bring Reinforcement-learning Into Radio Light Network for Massive Connections &THz, OWC, AI, 3D SLAM, D2D cell-free network, reinforcement learning  \\ \cline{2-4}		
& AI@EDGE& A secure and reusable Artificial Intelligence platform for Edge computing in beyond 5G Networks & AI for network automation, AI-enabled network applications, edge computing, security	\\ \cline{2-4}			
&DAEMON & Network intelligence for aDAptive and sElf-Learning MObile Networks 	& Network intelligence, AI, E2E architecture	\\ \cline{2-4}
& DEDICAT 6G & Dynamic coverage Extension and Distributed Intelligence for human Centric applications with assured security, privacy and trust: from 5G to 6G	 & Distributed intelligence, security and privacy, AI, blockchain, smart connectivity	\\ \cline{2-4}			
& Hexa-X & A flagship for B5G/6G vision and intelligent fabric of technology enablers connecting human, physical, and digital worlds	& High frequency, localization and sensing, connected intelligence, AI-driven air interface, 6G architecture	\\ \cline{2-4}			
& MARSAL & Machine learning-based, networking and computing infrastructure resource management of 5G and beyond intelligent networks	& Optical-wireless convergence, fixed-mobile convergence, distributed cell-free, O-RAN, AI, blockchain, secured multi-tenant slicing	\\ \cline{2-4}			
& REINDEER & REsilient INteractive applications through hyper Diversity in Energy Efficient RadioWeaves technology & Intelligent surfaces, cell-free wireless access, distributed radio, computing, and storage, channel measurement	\\ \cline{2-4}				
& RISE-6G & Reconfigurable Intelligent Sustainable Environments for 6G Wireless Networks& RIS, architecture and operation for multiple RISs, radio propagation modeling 	\\ \cline{2-4}			
& TeraFlow & Secured autonomic traffic management for a Tera of SDN flows & SDN, DLT, ML-based security, cloud-native arctecture\\ \hline
\end{tabular}
\end{table*}

Besides, many countries have announced and are implementing ambitious plans to launch 6G research and development initiatives. In Finland, the University of Oulu began ground-breaking 6G research as part of Academy of Finland’s flagship program \cite{Ref_WJ_Aazhang2019key} called 6G-Enabled Wireless Smart Society and Ecosystem (6Genesis), which focuses on several challenging research areas including reliable near-instant unlimited wireless connectivity, distributed computing and intelligence, as well as materials and antennas to be utilized in future for circuits and devices. As early as 2016, the U.S. Defense Advanced Research Projects Agency (DARPA), along with companies from the semiconductor and defense industries, has initiated the joint university microelectronic project (JUMP), among which the center for converged \si{\tera\hertz} communications and sensing seeks to develop technologies for a future cellular infrastructure. In March 2019, the US spectrum regulator -  the Federal Communications Commission (FCC) announced  to open up experimental licences for the use of frequencies between \SI{95}{\giga\hertz} and \SI{3}{\tera\hertz} for 6G and beyond. In October 2020, the Alliance for Telecommunications Industry Solutions (ATIS) announced the launch of the ``\textit{Next G Alliance}" \cite{Ref_WJ_nextGalliance}, an industry initiative that aims to advance North American mobile technology leadership in 6G over the next decade. Its ambition it to encompass the full lifecycle of 6G research and development, manufacturing, standardization, and market readiness. The founding members include AT\&T, T-Mobile,  Verizon, Qualcomm, Ericsson, Nokia, Apple, Samsung,  Google, Facebook, Microsoft, etc. In November 2019, China has officially kicked off the 6G technology research and development works coordinated by the Ministry of Science and Technology, together with five other ministries or national institutions. A promotion working group from government that is in charge of management and coordination, and an overall expert group that  is composed of 37 experts from universities, research institutes, and industry were established at this event. Later, it was announced that China aims to form 6G overall development ideas by the end of 2020. In November 2020, China launched what it claimed is the first 6G experimental satellite to test communications from space using high-frequency terahertz spectrum.
In early 2020, the Japanese government set up a dedicated panel including representatives from the private sector and academia to discuss technological development, potential use cases, and policy. Japan reportedly intends to dedicate around $\$2$ billion to encourage private-sector research and development for 6G technology. 
In late 2020, the government of South Korea has confirmed a plan to carry out a 6G trial in 2026 and is expected to spend approximately $\$169$ million over the course of five years to develop 6G technology. The trial aims to achieve \SI{1}{\tera\bps} in data transmission speeds and latency reduction to one-tenth of current 5G services. The government will initially push for tasks in six key areas (hyper-performance, hyper-bandwidth, hyper-precision, hyper-space, hyper-intelligence, and hyper-trust) to preemptively secure next-generation technology.

\section{Technological Enablers}\label{sec:enablers}
To well support disruptive use cases and applications, advanced technologies on transmission, networking, and computing would be developed and then applied in the 6G system.  This section provides a complete view of potential 6G technological enablers, which are categorized into several groups: \textit{new spectrum} consisting of mmWave, \si{\tera\hertz} communications, VLC, OWC, and DSM, \textit{new networking} that covers softwarization and virtualization, RAN slicing, O-RAN, and post-quantum security, \textit{new air interface} including massive MIMO, IRS, CoMP, cell-free massive MIMO, and new modulation techniques, \textit{new architecture} providing 3D coverage by means of integrating large-scale satellite constellation, HAP, and UAV with traditional terrestrial networks, and  \textit{new paradigm} empowered by the convergence of communication, computing, and storage resources, as well as the integration of AI, blockchain, digital  twin, and mobile networks. The principle, advantages, challenges, and open research issues for each identified technology are introduced.  

\begin{table*}[!hbtp]
	\centering
	\caption{Categorized key technology enablers with advantages and challenges}
	\label{tab:key_enablers}
	\scriptsize
	\begin{tabular}{|l|m{1cm}l|m{5.8cm}|m{6cm}|}
		\hline
		\textbf{Category}&\multicolumn{2}{l|}{\textbf{Enabler}}&\textbf{Advantages}&\textbf{Challenges}\\
		\hline\hline
		\multirow{5}{*}{Spectrum}&\multicolumn{2}{l|}{mmWave}&\multirow{2}{*}{high bandwidth, narrow beams, high integration level}&\multirow{2}{*}{severe attenuation \& blockage, low range}\\\cline{2-3}
		&\multicolumn{2}{l|}{THz}& &\\\cline{2-5}
		&\multicolumn{2}{l|}{OWC (including VLC)}&almost unlimited bandwidth, license-free, low cost, security, health-friendly&frail MIMO gain, HW implementation, noise, loss, nonlinearity, dispersion, pointing errors\\\cline{2-5}
		&\multicolumn{2}{l|}{DSM}&coexistence of licensed/unlicensed users&all-spectrum sensing, data processing \& management \\\hline
		\multirow{4}{*}{Networking}&\multirow{1}{*}{NFV \& SDN}&&high flexibility, low operational cost &service heterogeneity, SDN controller placement, auto network management and orchestration, E2E QoS control\\\cline{2-5}
		&\multirow{1}{*}{RAN Slicing}&&flexibility, resource efficiency, security&architectural framework supporting multi-use-case verticals  \\\cline{2-5}
		&\multirow{1}{*}{O-RAN}&&efficiency, intelligence, flexibility, dynamic&lack of tech. convergence \& standardization efforts\\\cline{2-5}
		&\multirow{1}{*}{Post-Quantum Security}&&service-based E2E security&AI \& ML deployment\\\hline
		\multirow{5}{*}{Air interface}&\multicolumn{2}{l|}{Massive MIMO}&high capacity, statistical multiplexing gain, high spectral efficiency, low expenditure, low energy consumption& extremely-large aperture, channel prediction, intelligent environment aware adaptation, holographic mMIMO, 6D positioning, large-scale MIMO radar\\\cline{2-5}
		&\multicolumn{2}{l|}{IRS, aka. RIS/SRE}&high MIMO gain, low implementation cost, low power~&rely on 3\textsuperscript{rd} party assessments, business framework\\\cline{2-5}
		&\multicolumn{2}{l|}{CoMP}&BS-level spatial diversity, ``cell-free'' potential&clustering, synchronization, channel estimation, backhaul \\\cline{2-5}
		&New&\multicolumn{1}{|l|}{Intell. OFDMA}&online MIMO precoding \& resource mapping&waveform design, out-of-band radiation  \\\cline{3-5}
		&modulation&\multicolumn{1}{|l|}{NOMA}&high power \& spec. efficiencies&specific D2D interface for cooperative decoding\\\hline
		\multirow{5}{*}{Architecture}&\multicolumn{2}{l|}{Large-scale LEO satellite}&ubiquitous coverage, resistance to natural disasters,&\multirow{2}{*}{integration with terrestrial networks, launching cost}\\
		&constellation&&lower channel loss and cost than GEO  &\\\cline{2-5}
		&HAP&& large coverage, unobstructed, flexible deployment, lower cost, easier access to infrastructure and better channel than&channel modelling, deployment, path planning, operational altitude, interference, energy limit, reliability\\\cline{2-3}\cline{5-5}
		&\multicolumn{2}{l|}{UAV}&satellite, new use scenarios&security, real-time demand \\\hline
		\multirow{13}{*}{Paradigm} & \multirow{4}{*}{AI}& \multicolumn{1}{|l|}{Deep learning}&automatic featuring, prediction & computational complexity \\\cline{3-5}
		&& \multicolumn{1}{|l|}{Federated learning}&protection to data privacy&communication overheads, heterogeneity \\\cline{3-5}
		&&\multicolumn{1}{|l|}{Transfer learning}& quick model adaptation and optimization at local&unification, gain measuring, dataset dissimilarity\\\cline{3-5}
		&&\multicolumn{1}{|l|}{AI as a service}&low latency AI service for end-user at terminals&new distributed AI techniques, new APIs\\\cline{2-5}
		&\multirow{1}{*}{Block-chain}&&immutability, decentralization, transparency, security and privacy &majority vulnerability, double-spending, transaction privacy leakage, scalability, quantum computing\\\cline{2-5}
		&\multirow{1}{*}{Digital twin}&&improved quality of products, services, processes, devices, etc. in Industry 4.0 and IoT&scalability, self-management, lack of models and methodologies, security and privacy\\\cline{2-5}
		&\multirow{1}{*}{Edge intelligence}&&resolves MEC issue caused by service requirement diversity among a massive number of users&customized AI algorithms, resource management and task scheduling \\\cline{2-5}
		&\multirow{1}{*}{CoCoCo convergence}&&resolves timeliness and resilience problems due to the coupling between communication, computation, and control systems &in-loop co-design methodologies \& frameworks\\\hline
	\end{tabular}
\end{table*}

\begin{table*}[!hbtp]
	\centering
	\caption{Categorized key technology enablers with impact on KPIs}
	\label{tab:enablers_kpis}
	\scriptsize
	\begin{tabular}{|l|ll|c|c|c|c|c|c|c|c|c|c|c|c|c|c|c|}
		\cline{4-18}
		\multicolumn{3}{l|}{}&\multicolumn{15}{c|}{\textbf{KPI}}\\\hline
		&&&\multirow{11}{*}{\rotatebox[origin=c]{90}{\parbox{33mm}{\raggedright Peak data rate}}}&\multirow{11}{*}{\rotatebox[origin=c]{90}{\parbox{33mm}{\raggedright User-experienced data rate}}}&\multirow{11}{*}{\rotatebox[origin=c]{90}{\parbox{33mm}{\raggedright Latency}}}&\multirow{11}{*}{\rotatebox[origin=c]{90}{\parbox{33mm}{\raggedright Mobility}}}&\multirow{11}{*}{\rotatebox[origin=c]{90}{\parbox{33mm}{\raggedright Connection density}}}&\multirow{11}{*}{\rotatebox[origin=c]{90}{\parbox{33mm}{\raggedright Network energy efficiency}}}&\multirow{11}{*}{\rotatebox[origin=c]{90}{\parbox{33mm}{\raggedright Peak spectral efficiency}}}&\multirow{11}{*}{\rotatebox[origin=c]{90}{\parbox{33mm}{\raggedright Area traffic capacity}}}&\multirow{11}{*}{\rotatebox[origin=c]{90}{\parbox{33mm}{\raggedright Reliability}}}&\multirow{11}{*}{\rotatebox[origin=c]{90}{\parbox{33mm}{\raggedright Signal bandwidth}}}&\multirow{11}{*}{\rotatebox[origin=c]{90}{\parbox{33mm}{\raggedright Positioning accuracy}}}&\multirow{11}{*}{\rotatebox[origin=c]{90}{\parbox{33mm}{\raggedright Coverage}}}&\multirow{11}{*}{\rotatebox[origin=c]{90}{\parbox{33mm}{\raggedright Timeliness}}}&\multirow{11}{*}{\rotatebox[origin=c]{90}{\parbox{33mm}{\raggedright Security and privacy}}}&\multirow{11}{*}{\rotatebox[origin=c]{90}{\parbox{33mm}{\raggedright CAPEX and OPEX}}}\\
		&&&&&&&&&&&&&&&&&\\
		&&&&&&&&&&&&&&&&&\\
		&&&&&&&&&&&&&&&&&\\
		&&&&&&&&&&&&&&&&&\\
		&&&&&&&&&&&&&&&&&\\
		\textbf{Category}&\textbf{Enabler}&&&&&&&&&&&&&&&&\\
		&&&&&&&&&&&&&&&&&\\
		&&&&&&&&&&&&&&&&&\\
		&&&&&&&&&&&&&&&&&\\
		&&&&&&&&&&&&&&&&&\\
		&&&&&&&&&&&&&&&&&\\
		\hline\hline
		\multirow{5}{*}{Spectrum}&\multicolumn{2}{l|}{mmWave}&$\medwhitestar$&$\medwhitestar$&$\checkmark$&&$\checkmark$&&&$\medwhitestar$&&$\medwhitestar$&$\medwhitestar$&&&&\\\cline{2-18}
		&\multicolumn{2}{l|}{THz}&$\medwhitestar$&$\medwhitestar$&$\checkmark$&&$\checkmark$&$\checkmark$&&$\medwhitestar$&&$\medwhitestar$&$\medwhitestar$&&&&\\\cline{2-18}
		&\multicolumn{2}{l|}{OWC (incl. VLC)}&$\medwhitestar$&$\medwhitestar$&$\checkmark$&&$\checkmark$&&&$\medwhitestar$&&$\medwhitestar$&$\medwhitestar$&&&&\\\cline{2-18}
		&\multicolumn{2}{l|}{DSM}&&$\checkmark$&&&$\medwhitestar$&$\checkmark$&&$\checkmark$&&&&&&&\\\hline
		\multirow{4}{*}{Networking}&\multicolumn{2}{l|}{NFV \& SDN}&&&&&&&&&$\checkmark$&&&&&$\checkmark$&$\medwhitestar$\\\cline{2-18}
		&\multicolumn{2}{l|}{RAN Slicing}&&$\checkmark$&$\checkmark$&&&&&&$\checkmark$&&&&&$\medwhitestar$&$\medwhitestar$\\\cline{2-18}
		&\multicolumn{2}{l|}{O-RAN}&&&&&&&&&&&&&&&$\medwhitestar$\\\cline{2-18}
		&\multicolumn{2}{l|}{Post-Quantum Security}&&&&&&&&&$\checkmark$&&&&&&$\medwhitestar$\\\hline
		\multirow{5}{*}{Air interface}&\multicolumn{2}{l|}{Massive MIMO}&$\medwhitestar$&$\medwhitestar$&&&&$\checkmark$&$\medwhitestar$&$\checkmark$&&&$\medwhitestar$&&&&\\\cline{2-18}
		&\multicolumn{2}{l|}{IRS, aka. RIS/SRE}&$\checkmark$&$\medwhitestar$&&&&$\medwhitestar$&$\checkmark$&$\checkmark$&&&$\medwhitestar$&&&&\\\cline{2-18}
		&\multicolumn{2}{l|}{CoMP}&$\checkmark$&$\checkmark$&&$\checkmark$&&$\checkmark$&$\checkmark$&&$\medwhitestar$&&$\checkmark$&$\checkmark$&&&\\\cline{2-18}
		&\multirow{2}{*}{New modulation}&\multicolumn{1}{|l|}{Intell. OFDMA}&&$\medwhitestar$&&&$\medwhitestar$&$\checkmark$&&&$\checkmark$&&&&&&\\\cline{3-18}
		&&\multicolumn{1}{|l|}{NOMA}&&$\medwhitestar$&&&$\checkmark$&$\medwhitestar$&$\medwhitestar$&$\medwhitestar$&$\checkmark$&&&$\checkmark$&&&\\\hline
		\multirow{3}{*}{Architecture}&\multicolumn{2}{l|}{Large-scale LEO satellite constellation}&&&&$\checkmark$&&&&&$\medwhitestar$&&&$\medwhitestar$&&&\\\cline{2-18}
		&HAP&&&&&$\checkmark$&&&&&$\medwhitestar$&&&$\medwhitestar$&&&\\\cline{2-18}
		&\multicolumn{2}{l|}{UAV}&&&&$\checkmark$&&&&&$\medwhitestar$&&&$\medwhitestar$&&&\\\hline
		\multirow{8}{*}{Paradigm} & \multirow{4}{*}{AI}& \multicolumn{1}{|l|}{Deep learning}&$\checkmark$&$\checkmark$&$\checkmark$&$\checkmark$&$\checkmark$&$\checkmark$&$\checkmark$&$\checkmark$&$\checkmark$&$\checkmark$&$\checkmark$&$\checkmark$&$\checkmark$&$\checkmark$&$\checkmark$\\\cline{3-18}
		&& \multicolumn{1}{|l|}{Federated learning}&&&&&&&&&&&&&&$\medwhitestar$&\\\cline{3-18}
		&&\multicolumn{1}{|l|}{Transfer learning}&&&$\medwhitestar$&&&&&&&&&&$\checkmark$&&\\\cline{3-18}
		&&\multicolumn{1}{|l|}{AI as a service}&&&&&&$\medwhitestar$&&&&&&&&&\\\cline{2-18}
		&\multicolumn{2}{l|}{Block-chain}&&&&&&&&&&&&&&$\medwhitestar$&\\\cline{2-18}
		&\multicolumn{2}{l|}{Digital twin}&&&&&&&&&$\checkmark$&&&&$\medwhitestar$&&\\\cline{2-18}
		&\multicolumn{2}{l|}{Edge intelligence}&&$\checkmark$&$\medwhitestar$&&$\medwhitestar$&$\checkmark$&&&$\checkmark$&&&&$\checkmark$&&\\\cline{2-18}
		&\multicolumn{2}{l|}{CoCoCo convergence}&&&$\medwhitestar$&&&&&&$\medwhitestar$&&&&$\medwhitestar$&&\\\hline\hline
		\multirow{2}{*}{\textbf{Legend}}&\multicolumn{17}{l|}{$\checkmark$:\quad Generic/weak impact}\\
		&\multicolumn{17}{l|}{$\medwhitestar$:\quad Specialized/critical impact}\\\hline
	\end{tabular}
\end{table*}

\subsection{New Spectrum} \label{subsec:newspectrum}
Next generation cellular networks provide a good capability of heterogeneous radio access technology (RAT), where the legacy RAT with low radio frequencies and the line-of-sight (LOS)-dependent RATs (\si{\tera\hertz}, VLC, and OWC) can co-exist well. \si{\tera\hertz}, VLC, and OWC may construct a new layer in the hierarchical RAN architecture (e.g., picocells), where heterogeneous cells with different RAT are overlaying on each other. The ideology is similar to the introduction of mmWave in the 5G networks.

\subsubsection{Millimeter Wave}
The mmWave technology has been introduced by the 5G new radio, and believed to remain as an essential component in future 6G networks. Compared to legacy RF technologies working below \SI{6}{\giga\hertz}, it significantly broadens the available bandwidth with new carrier frequencies up to \SI{300}{\giga\hertz}. Such a huge new bandwidth, as Shannon's theorem reveals, will inflate the radio channel capacity and quench the imminent thirsty for higher data rate. Meanwhile, the shorter wavelength also leads to smaller antenna size. This not only improves the portability and integration level of device, but also allows to increase the dimension of antenna arrays and therewith narrow the beams \cite{Ref_BH_wang2018millimeter}, which is beneficial to specific applications such like detection radars and physical layer security. Furthermore, the atmospheric and molecular absorption exhibit highly variant characteristics at different frequencies across the mmWave band, providing potential for diverse use cases. On the one hand, low attenuation can be observed at some special bands such as \SI{35}{\giga\hertz}, \SI{94}{\giga\hertz}, \SI{140}{\giga\hertz}, and \SI{220}{\giga\hertz}, making long-distance peer-to-peer communications possible at these frequencies; on the other hand, severe propagation loss is experienced at some ``attenuation peaks'' such as \SI{60}{\giga\hertz}, \SI{120}{\giga\hertz}, and \SI{180}{\giga\hertz}, which can be exploited by short-range covert networks with stringent safety requirements \cite{Ref_BH_wang2018millimeter,Ref_BH_ippolito1981radio}. Currently, the standardization efforts in the mmWave field are mainly focusing on the \SI{60}{\giga\hertz} band for indoor use, e.g. the ECMA-387~\cite{Ref_BH_ajorloo2013modeling}, the IEEE 802.15.3c~\cite{Ref_BH_ieee80215tg3c}, and the IEEE 802.11ad~\cite{Ref_BH_nitsche2014ieee}.

Accompanying to the benefits of mmWave technologies, there also come new challenges. First of all, the broad bandwidth in mmWave band and high transmission power can lead to severe non-linear signal distortions, which proposes higher technical requirements for the integrated circuits than those for RF devices. Meanwhile, since the effective transmission range of mmWave, especially in the \SI{60}{\giga\hertz} band, is severely limited by the atmospheric and molecular absorption, mmWave channels are commonly dominated by the LOS path. This becomes a major drawback that is further magnified by the poor diffraction at this short wavelength, which causes a strong blockage loss in scenarios with dense presence of small-scale obstacles such as vehicles, pedestrians, or even the human body of user itself~\cite{Ref_BH_han20173d}. The high propagation loss and LOS-dependency also significantly raises the channel state sensitivity to the mobility, i.e., the impact of fading is much stronger than that in the RF bands. The demand for an outstanding mobility management becomes therefore unprecedentedly high. Furthermore, in scenarios with dense links coexisting, especially for indoor environments, the interference among different access points will be significant, interference management approaches are therefore called for ~\cite{Ref_BH_al2018survey}.

\subsubsection{Terahertz Communications}
Despite of its current abundance in spectral redundancy, mmWave is hardly adequate to tackle down the increasing cravenness on bandwidth for another decade. Looking forward to the 6G era, wireless technologies operating at even higher frequencies, such as \si{\tera\hertz} or optical frequency bands, are expected to play an important role in the next generation RAN, providing extremely high bandwidth.

Similar to mmWave, \si{\tera\hertz} waves also suffer from high path loss and therefore highly rely on directive antennas and LOS channels while providing a very limited coverage. However, when a satisfactory LOS link is available, the high carrier frequency brings a bandwidth that is significantly higher than any legacy technology, which makes it possible to simultaneously provide ultra-high performance in aspects of throughput, latency, and reliability. Moreover, compared to both mmWave systems working at lower frequencies and wireless optical systems working in higher frequency bands, \si{\tera\hertz} communication systems are insensitive to atmospheric effects, which eases the tasks of beamforming and beam tracking. This shapes \si{\tera\hertz} communication into a good supplementary solution in addition to the mainstream RF technologies for specific use cases, such as indoor communications and wireless backhaul; and a competitive option for future cyber-physical applications with extreme QoS requirements, such like real-time VR/AR~\cite{Ref_BH_chen2019survey}.

Furthermore, the high carrier frequency also allows smaller antenna size for higher integration level. It is expected \cite{Ref_BH_zhang20196g} that over 10~000 antennas can be embedded into a single \si{\tera\hertz} BS to provide hundreds of super-narrow beams simultaneously, so as to overcome the high propagation loss, and to simultaneously achieve extremely high traffic capacity together with massive connectivity, which assemble to unlock its applications in ultra-massive machine-type communications such as Internet-of-Everything (IoE) \cite{Ref_MAH_Han2017MTC}.

Nevertheless, while \si{\tera\hertz} outperforms mmWave in many ways, it also faces stronger technical challenges, especially from the aspect of implementing essential hardware circuits, including antennas~\cite{Ref_BH_vettikalladi2019sub}, amplifiers~\cite{Ref_BH_tucek2016operation}, and modulators. Especially, it has been since long the most critical challenge for practical deployment of \si{\tera\hertz} technologies, to efficiently modulate baseband signals onto such high frequency carriers with integrated circuits. To address this issue, a great effort has been made over the past decade, leading to a prosperous set of developments, most of which are solid state \si{\tera\hertz} systems that rely on frequency mixing, such as the one reported in \cite{Ref_BH_koenig2013wireless}. Recently, it has also been discussed to apply spatial direct modulation in \si{\tera\hertz} systems, so as to directly modulate baseband signals to \si{\tera\hertz} band without any intermediate frequency stage.

\subsubsection{Visible Light Communications} \label{subsubsec:vlc}
VLC works in the frequency range of \SIrange{400}{800}{\tera\hertz}. Differing from the RF technologies in lower \si{\tera\hertz} range that use antennas, VLC relies on illumination sources -- especially light-emitting diodes (LEDs) -- and image-sensor or photodiode arrays to implement the transceivers. With these transceivers, a high bandwidth can be easily achieved with low power consumption (\SI{100}{\milli\watt} for \SIrange{10}{100}{\mega\bps}) without generating electromagnetic or radio interference  \cite{Ref_BH_sevincer2013lightnets}. The good power efficiency, the long lifetime (up to 10 years) and low cost of mainstream LEDs, in addition to the unlicensed access to spectrum, makes VLC an attractive solution for use cases sensitive to battery life and access cost, such like massive IoT and wireless sensor network (WSN). Moreover, VLC also exhibits better propagation performance than RF technologies do in some non-terrestrial scenarios, such as aerospace and underwater, which can be important part of the future 6G ecosystem, as we will introduce later in Sec.~\ref{subsec:new_arch}.

Compared to RF, the MIMO gain in VLC, especially in indoor scenarios, is very frail. This roots in the high coherence among the propagation paths, i.e. the low spatial diversity. Though this coherence can be somehow reduced by using spaced LED arrays \cite{Ref_BH_al2018optical}, MIMO-VLC is also challenged by the design and implementation of receivers: non-imaging receivers are extremely sensitive to their spatial alignments with the transmitters, while imaging receivers are not applicable in cost-critical use cases for their high prices~\cite{Ref_BH_pathak2015visible}. Therefore, there has been so far no MIMO method standardized into the mainstream VLC physical layer of IEEE 802.15.7, despite of the persistent efforts made in academia since a decade~\cite{Ref_BH_zeng2009high,Ref_BH_nuwanpriya2015indoor,Ref_BH_huang2018transceiver}. Therefore, the beamforming in VLC, differing from the MIMO-based RF beamforming, is implemented by a special optical device known as spatial light modulator (SLM)~\cite{Ref_BH_wu2014transmit}.

Similar to mmWave and \si{\tera\hertz} technologies, VLC also relies on LOS channels, since it has neither ability to penetrate, nor sufficient diffraction to bypass common obstacles. Meanwhile, due to the concerns about adjacent cell interference and almost ubiquitous environment light noise, VLC systems generally require directive antennas with narrow beams.  These facts make VLC systems highly sensitive to the position and mobility of users, leading to a high requirement of beam tracking. On the other hand, this feature can be also exploited for advantages in certain use scenarios, such like better accuracy in indoor positioning~\cite{Ref_BH_zhuang2018survey}, and lower interference in vehicular communications~\cite{Ref_BH_memedi2020vehicular}.

Another key technical challenge for VLC roots in the open and unregulated (to be more specific, \emph{unregulatable}) access to visible light spectrum, which implies a higher security risk and calls for a more stringent security requirement in VLC systems, when compared to legacy cellular systems in licensed RF bands. Regarding this, physical layer security has been extensively investigated as a promising solution \cite{Ref_BH_arfaoui2020physical}.

\subsubsection{Optical Wireless Communications} 
OWC points to wireless communications \cite{Ref_WJ_elgala2011indoor} that use infrared (IR), visible light, or ultraviolet (UV) as transmission medium. It is a promising complementary technology for traditional wireless communications operating over RF bands. OWC systems operating in the visible band are commonly referred to as VLC, which attracts much attention recently and is discussed separately in Sec.~\ref{subsubsec:vlc}. The optical band can provide almost unlimited bandwidth without the permission from the regulators worldwide. It can be applied to realize high-speed access at low cost thanks to the availability of optical emitters and detectors. Since the IR and UV waves have a similar behavior as the visible light, the security risks and interference can be significantly confined and the concern about the radio radiation to human health can be eliminated. It is expected to have obvious advantages in deployment scenarios such as vehicular communications in smart transportation systems, airplane passenger lighting,  and medical machines that are sensitive to electromagnetic interference. Despite its advantage, OWC suffers from the impairments such as ambient light noise, atmospheric loss, nonlinearity of LEDs, multi-path dispersion, and pointing errors \cite{Ref_WJ_kahn1997wireless}. 

In OWC,  LED or laser diode (LD) is applied to convert an electrical signal to an optical signal at the transmitter and the receiver uses a photodiode (PD) to convert the optical signal into electrical current. The information is conveyed by modulating the intensity of optical pulse simply through widely-used scheme such as on-off keying or pulse-position modulation, as well as advanced multi-carrier schemes \cite{Ref_WJ_ohtsuki2003multi} such as OFDM to get higher transmission rate. To support multiple users in a single optical access point, OWC can apply not only typical electrical multiplexing technologies such as time-division, frequency-division, and code-division multi-access, but also optical multiplexing such as wavelength-division multi-access. Optical MIMO technology \cite{Ref_WJ_zeng2009high} is also implemented in OWC, where multiple LEDs and multiple PDs are applied, in contrast to the multiple antennas in a typical MIMO system operating in the RF band.   

The optical system applying image sensors to detect the optical pulse also called optical camera system. The imagine sensor can convert the optical signal into the electrical signal, which has the advantage of easier implementation due to the wide spread of camera-embedded smart phones. On the other hand, terrestrial point-to-point OWC also known as free-space optical (FSO) communications \cite{Ref_WJ_juarez2006free} take place at the near IR band. Using high-power high-concentrated laser beam at the transmitter, the FSO system can achieve high data rate, i.e., \SI{10}{\giga\bps} per wavelength, over long-distance (up to \SI{10~000}{\kilo\meter}). It offers a cost-effective solution for the backhaul bottleneck in terrestrial networks, enables crosslinks among space, air, and groud platforms, and facilitates high-capacity inter-satellite links for the emerging LEO satellite constellation. 
Furthermore, there has also been a growing interest on ultraviolet communication \cite{Ref_WJ_xu2008ultraviolet} as a result of recent progress in solid state optical transmitter and detector for non-LOS UV communications that offer broad coverage and high security.

\subsubsection{Dynamic Spectrum Management}
Alongside the continuous mining of unused spectrum at ever-higher frequencies, there is a second approach towards our vision of bandwidth prosperity in 6G: to improve the radio resource utilization rate by DSM. 

The idea of DSM dates back to the well-known listen-before-talk (LBT) protocol applied in IEEE 802.11, which treats all users equally in a contention-based control of access to the spectrum, and one can exploit a band only when it is not occupied by the others. In the unlicensed industrial, scientific and medical (ISM) band, LBT has demonstrated a great success in collision and interference control. Meanwhile, in the licensed spectrum, as reported by the FCC of the U.S., it is the under-utilization of spectrum caused by regulated access ``\emph{a more significant problem than the physical scarcity of spectrum}''~\cite{Ref_BH_marcus2002federal}. This fact has raised intense research interest on the topic of LBT-like dynamic spectrum sharing among various systems with heterogeneous RATs and different priorities to access licensed/unlicensed bands. Driven by the successful development of software-defined radio technologies, these research efforts gave birth to the technology of cognitive radio (CR)~\cite{Ref_BH_haykin2005cognitive}, which became rapidly mature in the first decade of this century. Since the LTE era, it has become a distinguished topic in the field of wireless networks to study the DSM in coexistence of licensed cellular systems and unlicensed ISM-band technologies~\cite{Ref_BH_voicu2016inter}. 

Regarding the future 6G systems, the demand for DSM is becoming unprecedentedly imperative. On the one hand, radio access to ISM bands (especially the IEEE 802.11 bands) is becoming nowadays almost \textcolor{red}{a} standard functionality of mainstream cellular terminals, making it a universal solution to provide extra network capacity in scenarios with dense users. On the other hand, since it is impossible to reserve the broad bands of 6G new spectrum for licensed use only (saliently the visible light spectrum), 6G system is foreseen to suffer in the unlicensed part of its spectrum from the ubiquitous existence of interference by other systems and environmental noises, which are highly dynamic and environment-dependent. 6G systems therefore must be able to dynamically and cognitively select the most appropriate operating band with respect to the instantaneous situation.

There are numerous technical challenges that 6G DSM is facing. In perspective of the hardware implementation, it is mainly troubled by the broadness of 6G new spectrum that leads to a difficulty in designing transceivers capable of dynamic all-spectrum sensing~\cite{Ref_BH_akyildiz20206g}. The 6G front-end must be capable to carry out a rapid and power-efficient spectrum sensing across the ultra-wide 6G bands, so as to enable an online radio environment cognition and a timely adaptation to spectrum access. On the system level, for better efficiency and safety of the DSM, the physical layer CR based on spectrum sensing shall be further completed by an awareness of cyber-physical level context information, in order to obtain a deeper understanding in the communication environment, including terrain scenario, traffic pattern, local regulation, etc. This leads to challenges in various aspects of context-awareness, ranging from data provisioning to data ownership \cite{Ref_BH_kliks2020beyond}.

\subsection{New Networking} 
To deploy the aforementioned use cases utilizing the discussed key enabling technologies, the network infrastructure is expected to be flexible, intelligent, and open to multi-vendor equipment and multi-tenancy. To that aim, softwarization and virtualization of 6G network is the first and foremost step. Based on such architecture, the network slicing and resource isolation in both core and RAN domains are considered as novel mechanisms to decrease complexity in the management and orchestration operations. These aspects, along with their privacy and security concerns, of 6G mobile networks are discussed in this subsection.

\subsubsection{Softwarization and Virtualization } 
Most of the network functions of the 5G core network (5GC) and next-generation RAN (NG-RAN) architecture are virtualized using network function virtualization (NFV) technology, which provides high flexibility in order to adapt to various scenarios, requirements, and use cases of next generation communication systems \cite{Ref_MAH_Habibi2019Survey}. In practice, however, NFV faces some critical challenges that are required to be solved, such as the increased amount of virtual network functions (VNFs), the various requirements of the different tenants, the resources orchestration of the VNFs in a shared infrastructure, the complexity in management and orchestration operations, and others \cite{Ref_MAH_Laghrissi2019VNF}. Therefore, the allocation of virtual network resources, the management and orchestration of the VNFs in a multi-tenant environment require cutting-edge tools and promising solutions to be addressed, such as the AI techniques \cite{Ref_WJ_jiang2017experimental} and ML algorithms \cite{Ref_WJ_jiang2017intelligent}. In this context, the European Telecommunication Standards Institute (ETSI) Industry Specification Group (ISG) introduced the Experiential Network Intelligence (ENI) working group in order to improve the experience of network operators and add value to the teleco provided services \cite{Ref_MAH_Wang2018NFV}. The main objectives of the ENI is to exploit AI and ML techniques in order to adjust the VNFs of the networked services based on dynamic changes in the requirements of the end users, the conditions of the environments, and the goals of the business. The current specification of the ENI is in its initial phase. Further work is needed to explore the deployment of the AI and ML in the NFV of the beyond 5G and 6G mobile networks.   

The software-defined networking (SDN) is considered as one of the most critical key enablers of 5G mobile networks, integrated with NFV capabilities, they are able to offer high flexibility in network management and achieve high effectiveness in service modularity \cite{Ref_MAH_Habibi2019Survey}. Based on its integral role and performance in the 5G mobile network, the SDN is going to be indispensable for the evolution of beyond 5G and 6G mobile networks as well and will continue to play a crucial role within their management, orchestration, architectures, etc. \cite{Ref_WJ_Letaief2019roadmap}. Despite its numerous advantages in 5G, the SDN technology has a number of key research problems that are needed to be addressed in order to thwart full exploitation of its potentials in the future generation of communication networks. These key challenges that continue to confront SDN in the wake of 6G include, but not limited to: effectively resolving the optimal placement of the SDN controller in the future networks \cite{Ref_MAH_Ksentini2016SDN, Ref_MAH_Alevizaki2020SDN}, efficiently maintaining the end-to-end and timely global views of the dynamic network topology and its links states \cite{Ref_MAH_Khan2017SDN}, exploiting AI/ML techniques for automated network management \cite{Ref_WJ_jiang2017autonomic} and orchestration \cite{Ref_MAH_Chemouil2019SDN}, and traffic engineering with guaranteed stringent QoS requirements of heterogeneous services \cite{Ref_MAH_Tomovic2019SDN}.

\subsubsection{RAN Slicing}
Slicing the RAN architecture, using SDN and NFV technologies, is an emerging research direction towards cloudification, virtualization, and centralization of RAN resources in beyond 5G and 6G mobile networks \cite{Ref_MAH_Elayoubi2019RANSlicing}. The slicing-aware NG-RAN architecture helps mobile operators to efficiently slice the entire infrastructure (or a part of it) based on the requirements of end users and vertical industries \cite{Ref_MAH_Habibi2018Slicing}. Such a classification is also applicable in terms of slicing the NG-RAN into eMBB, URLLC, and mMTC subnets. The resources required by these three types of RAN slices are divided into physical and virtual resources. The physical resources are managed by the 3GPP network slicing management system \cite{Ref_MAH_3GPP2018RANSlicing} whereas the virtual resources are managed by the ETSI network function virtualization-management and orchestration (NFV-MANO) \cite{Ref_MAH_ETSI2017RANSlicing}. 

Towards an efficient RAN slicing, currently, some of the radio processing functionalities of a next-generation NodeB (gNB) in the NG-RAN are accommodated as VNFs \cite{Ref_MAH_Hirayama2019RANSlicing}, namely the centralized unit (CU) and distributed unit (DU), and some others are distributed as physical network function (PNF), namely the radio unit (RU). The VNFs run on points of presence (PoPs) \cite{Ref_MAH_ETSI2015RANSlicing}, while the PNFs are implemented on dedicated hardware in the cellular network sites. In the next generation of mobile networks, the full virtualization of CU and DU, and the partial/full virtualization of RU, can lead to a number of advantages, such as increased performance of the RAN architecture, deployment of RAN slice subnets, decreased network expenditure, simplified operations and management of the network, etc. \cite{Ref_MAH_Habibi2020SlicingMultiple}. We, thus, firmly believe that further studies are needed to address the virtualization of RU functionalities towards a virtualized and slicing-aware RAN of the 6G mobile networks \cite{Ref_MAH_Habibi2020SlicingChapter}.

Focusing on the function split in the NG-RAN, the current distribution of the radio processing functions over gNB components is executed statically without taking into consideration the type of RAN slices. When supporting a large amount RAN slices in types of eMBB, URLLC, and mMTC, the one-size-fits-all function splitting architecture in NG-RAN is not efficient in terms of resource allocation and network performance \cite{Ref_MAH_Hirayama2019RANSlicing}. We strongly believe that a customized and dynamic distribution of the gNB functionalities is needed in order to satisfy the service requirements of the three aforementioned types of RAN slices in future networks. Such a dynamic distribution of gNB functionalities shall improve the utilization of physical and virtual resources, enhance the NG-RAN performance, and maintain a significant level of isolation and customization among RAN slices of different types while considering the metrics of service level agreements between the mobile operator and tenants \cite{Ref_MAH_Habibi2018SLA}. 

The NG-RAN is expected to support a massive amount of RAN slice subnets. Each type of RAN slice subnet fulfills the service requirements of a singe type of use case of a single tenant. Nevertheless, there is a large number of vertical industries -- such as automobile, manufacturing, power grid, and others -- which are consisting of multiple use cases \cite{Ref_MAH_Habibi2020SlicingMultiple}. Providing RAN slice subnets for multiple use cases consisting of vertical industries is a crucial research problem that is needed to be addressed in the next generation of mobile networks. One of the crucial parts of this research problem is to design an extensive architectural framework for the 6G RAN domain in order to support the RAN slice subnets of multi-use-case verticals. This management and orchestration framework shall effectively manage the CU, DU, and RU of per-vertical per-use-case RAN slices in 6G RAN using AI and ML. 

\subsubsection{Open-RAN} 
The key concepts of O-RAN, including its vision, architecture, interfaces, technologies, objectives, and other important aspects, were introduced for the first time by the O-RAN alliance in a white paper \cite{Ref_MAH_ORAN2018ORANWhitePaper}. The O-RAN alliance has then further studied the use cases leveraging the O-RAN architecture to demonstrate its capability in real time behavior in \cite{Ref_MAH_ORAN2020ORANWhitePaper}. The main objective of the openness and intelligence in RAN architecture is to build a radio network that is resource efficient, cost effective, software-driven, virtualized, slicing-aware, centralized, open source, open hardware, intelligent,  and therefore more flexible and dynamic than any previous generation of mobile networks. To do so, the research community has introduced the utilization of AI and ML techniques on every single layer of the RAN architecture to fulfill the requirements of dense network edge in beyond 5G and 6G mobile communication systems. 

By opening up the RAN from a singe vendor environment to a standardized, open, multi-vendor, and ML and AI powered hierarchical controller structure, it allows third parties and mobile operators to dynamically deploy innovative applications and emerging services that cannot be deployed or supported in legacy RAN architectures. In addition, the O-RAN is built upon the NFV-MANO reference architecture proposed by ETSI, which deploys commercial off-the-shelf hardware components, virtualization techniques, and software pieces. The virtual machines abstracted (or virtualized) from the underlying physical resources are easily created, deployed, configured, and decommissioned. Such a virtualized environment and virtual resources, therefore, do not only bring flexibility to the O-RAN architecture, but also decrease the CAPEX/OPEX and energy consumption towards 6G communication networks. 

Despite the flexibility and interoperability that O-RAN offers to mobile operators, it also has a number of key problems that require further research efforts for its full realization in future mobile networks, such as the convergence of different vendors' technologies and various operators on the same platform, the harmonization of different management and orchestration frameworks to deliver enhanced QoE, lacked standards for the validation and troubleshooting of performance issues related to the network, and others. To overcome these challenges, researchers from industry and academia are also expected to take part in theoretical analysis and practical roll-outs of this technology towards an open and intelligent RAN for 6G mobile networks.

\subsubsection{Post-Quantum Security}
In 5G communication systems, security and privacy are considered as the most critical components for business continuity. This issue has even been raised to the stage of international politics, for example, some countries are proposing sanctions to ban 5G  hardware and software from certain vendors, claiming ``to protect their networks and citizens''. Focusing on the technical aspect of the 5G security, together with user equipment, the rise of new business and innovations such as vertical industries, mobile virtual network operators, and multi-tenancy, have put a burden on the commercial deployment of 5G networks. For instance, mMTC types of communication services require lightweight security components while eMBB and URLLC types of communication services demand high efficient security schemes. Another example is multi-tenant aspect, which implies the lack of a central authentication server so that the subscribers’ identities must be confirmed in either a decentralised or a collaborative fashion. It is envisioned that 6G systems will encounter more challenging security problems over current 5G systems. Many cutting-edge mechanisms are under exploration to meet the requirement of high security and privacy in the next-generation networks, such as E2E security, distributed authentication, and deep learning-based intrusion detection.
To ensure the E2E security for 6G, the deployment of the AI techniques can play a significant role in the design, implementation, and optimization of security protocols in order to protect network, user equipment, and vertical industries from unauthorized access and threats. To allow for trustworthy among communication participants, distributed authentication relevant to users authenticating against gNB, and among network functions (RU,
DU, and CU), will be designed by leveraging blockchain technology. 

In addition to conventional threats, the rise of quantum computing imposes a big challenge on network security. The cryptosystems used today can be divided into two categories: symmetric and asymmetric \cite{Ref_WJ_cheng2017securing}. The former shares a common secret key between two communicating parties, and a message is encrypted at the sender and is decrypted at the receiver using this key. An example of symmetric cryptography that is widely used today is advanced encryption standard (AES), which has been standardized in 2001 by National Institute of Standards and Technology (NIST) for confidentiality and integrity. The process of an exhaustive search for all possible keys can be substantially accelerated by Grover's algorithm, making such cryptosystems insecure once quantum computers come. In an asymmetric cryptosystem,  a public key is applied to encrypt messages by anybody while only the owner of the corresponding private key can decrypt cipher-texts. The public key schemes are also employed for the implementation of digital signature, where a signature is generated from the private key while everyone can utilize the public key to validate this signature. Current asymmetric cryptography schemes such as Rivest-Shamir-Adleman (RSA), elliptic curve cryptosystem (ECC), and digital signature algorithm (DSA) are based on the hardness of solving some number-theoretic problems such as  integer factorization and discrete logarithms. However, mathematician Peter Shor \cite{Ref_WJ_monz2016realization} revealed that quantum computers can solve such problems efficiently, which makes all these public-key schemes completely broken. 

It is still unclear when practical quantum computers will be available, but recent advances in quantum technologies demonstrate the urgency of exploring post-quantum  security for communication networks. Since 6G networks will be deployed around 2030 and will last for several decades, long-term threats arising from quantum computing could be seriously taken into account during the design and implementation of 6G systems. Consequently, the research and development of quantum-resistant cryptographic algorithms and technologies, also called post-quantum cryptography, which are secure against both quantum and classical computers play a vital role for the success of 6G. According to the initial recommendation of NIST \cite{Ref_WJ_nist2016report}, the cryptographic schemes based on lattices, codes, hash, and multivariate polynomials could be usable in the quantum era. 

Moreover, another quantum technology called quantum communication \cite{Ref_WJ_pan2020experimental} can significantly enhance the security level of data transmission, which might have application potential in 6G networks. In terms of the laws of quantum physics, if an adversary eavesdropper measures the state of superposition from particles, typically photons of light for transmitting data along optical fibers, their super-fragile quantum state “collapses” to either 1 or 0. Since a single particle carrying qubits is inseparable, the eavesdropper cannot replicate it. It means  that the eavesdropper can’t tamper with the qubits without leaving behind a telltale sign of the activity. In theory, quantum communication could provide absolute security and offer new solutions to a high level of security that traditional communication systems are unable to implement \cite{Ref_WJ_liao2020long}. 

\subsection{New Air Interface} 
The union of OFDM and small-scale active MIMO antenna arrays has ruled the world of cellular radio access network over the entire era of 4G, and is still continuously showing its dominance in recent progresses of 5G. However, the grandiose march towards ever-higher carrier frequency, as we have seen in Sec.~\ref{subsec:newspectrum}, is squeezing the last drop of technical potential from this air interface. Facing new challenges such as high propagation loss and low NLOS path diversity, various emerging technologies are expected to play their key roles in the evolution towards the next generation air interface, which shall be capable to fully exploit the advantages of 6G new spectrum and support future use cases with extreme performance requirements. This evolution is primarily expected to make several significant shifts and extensions by means of MIMO: from small-scale to massive, from active antenna to passive reflective surface, from physical layer to network layer. New modulation and multiplexing schemes, as complements, are also arising on the horizon of the next decade.

\subsubsection{Massive MIMO}
In legacy cellular networks the MIMO is divided into two types: the point-to-point MIMO and the multi-user MIMO. In the former, multiple antennas are installed in both user equipment and base station, however a single user equipment is served at a single time. In the latter, an antenna array is installed in the base station which provides connection to many user equipment under its respective coverage area. In order to further enhance user experience, increase throughput, and scale up the statistical multiplexing gain, the concept of massive MIMO was introduced in order to address the shortcomings of conventional multi-user MIMO \cite{Ref_MAH_Habibi2019Survey}. Since then, the massive MIMO has been considering one of the key enablers of the legacy wireless communication systems. In addition, the massive MIMO is also expected to provide a significant increase in the system capacity, higher statistical multiplexing gain, spectral efficiency, lower CAPEX/OPEX, decreased energy consumption, and many other advantages in beyond 5G and 6G cellular networks.

To commercially deploy massive MIMO, many operators have configured base stations with 64 fully digital transceiver chains which have proved its realization in the 5G mobile networks \cite{Ref_MAH_Bjornson2019MIMO}. These rollouts proved that limitations due to pilot contamination have addressed and for a better spectral efficiency the relevant signal processing methods have been developed and deployed. To pave the way towards the realization of massive MIMO in beyond 5G and 6G mobile networks, the authors in \cite{Ref_MAH_Bjornson2019MIMO} and \cite{Ref_MAH_Huang2020MIMO} have outlined a number of research challenges including the \textit{(i)} deployment  of extremely large aperture arrays; \textit{(ii)} limitation in channel prediction; \textit{(iii)} implementation of intelligent environment aware adaptation; \textit{(iv)} fundamental limits of wireless communication with holographic Massive MIMO; \textit{(v)} six-dimensional positioning; and \textit{(vi)} large-scale MIMO radar. These research problems give a chance to researchers working in academia, industry, and standardization organizations to concentrate their attention towards further improvement in the realization of massive MIMO in next generation of communication networks.  

\subsubsection{Intelligent Reflecting Surfaces}
While releasing a significant bandwidth to support high throughput, the use of high frequency bands over \SI{10}{\giga\hertz} also introduces new challenges, such as higher propagation loss, lower diffraction and more blockage. In the frequency range of mmWave, massive MIMO has been proven effective in realizing active beamforming to provide high antenna gain for overcoming the channel loss. Nevertheless, its capability can be insufficient for the future 6G new spectrum, as we have discussed earlier in Sec.~\ref{subsubsec:vlc}. Among all potential candidate solutions to enhance current beamforming approaches, the technology of IRS has been widely considered promising for 6G mobile networks.

The so-called IRS, a.k.a. reconfigurable intelligent surfaces (RIS)~\cite{Ref_BH_mossallamy2020reconfigurable}, is assembled by a category of programmable and reconfigurable  material sheets that are capable of adaptively modifying their radio reflecting characteristics. When attached to environmental surfaces, e.g. walls, glass, ceilings, etc., IRS enables to convert parts of the wireless environment into smart reconfigurable reflectors, known as smart radio environment (SRE)~\cite{Ref_BH_renzo2020smart}, and therewith to exploit them for a passive beamforming that can significantly improve the channel gain, at low costs of implementation and power consumption in comparison to active massive MIMO antenna arrays. Moreover, unlike antenna arrays that must be compact enough for integration, SREs are implemented on large-size surfaces apart from the UEs, making it easier for them to realize accurate beamforming with ultra-narrow beams, which are essential for some applications such like physical layer security. Furthermore, unlike active mMIMO antenna arrays that must be specifically implemented for every individual RAT, the passive reflection mechanism that IRS is relying on works almost universal for all RF and optical frequencies, which is especially cost beneficial for the 6G systems that work in an ultra broad spectrum.

Though IRS is showing a great technical competitiveness in context of the 6G new spectrum, it still lacks mature techniques for accurate modeling and estimation of the channels and the surface themselves, especially in the near-field range. Moreover, a commercial deployment is only possible after addressing the business concern, that IRS relies on external assessments such like buildings that do not belong to the MNOs. Therefore, it calls for a thoughtful design and standardization of framework providing essential interfaces, agreements, and signaling protocols, so that 6G operators become capable to widely access and exploit IRS-equipped objects in public and private domains.

\subsubsection{Coordinated Multi-Point and Cell-Free}
CoMP refers to a class of technologies that allow multiple access points to jointly serve multiple mobile stations, so that a network layer MIMO can be realized to increase the spatial diversity on the top of classical physical layer MIMO approaches. Therefore, it is also known as network MIMO or cooperative MIMO. CoMP was initially introduced by 3GPP in its Release 11~\cite{Ref_BH_3gpp2013coordinated} for LTE Advanced systems. With recent evidences of its potentials in mitigating downlink inter-cell interference and joint user detection in uplink, CoMP is expected to play an important role in 5G~\cite{Ref_BH_li20145g}. In the upcoming 6G era, regarding the new spectrum over \SI{10}{\giga\hertz}, the CoMP technologies that make use of base station level diversity will become an important complement to the traditional antenna-level spatial diversity, as the latter can be minimized by the dense blockage phenomena in high-frequency bands.

Furthermore, since CoMP is generally suggesting every UE to simultaneously hold multiple links to different access points (even when they are of the same RAT), it reveals the feasibility of a novel ``cell-free'' RAN architecture, where numerous single-antenna access points distributed over the coverage area are connected to a central processing unit, and jointly serving all UEs by coherent transmission in a CoMP fashion~\cite{Ref_BH_ngo2017cell}. Recent study has shown that such cell-free massive MIMO are able to outperform traditional cellular massive MIMO while also reducing the fronthaul signaling~\cite{Ref_BH_bjornson2019cell}.

As a trade-off for the performance gain originated from its nature of cooperative decoding, CoMP also has to face some key technical challenges caused by the same reason. First of all, the performance of CoMP highly relies on the clustering of cooperating base stations, so an appropriate clustering scheme must be found, which has been a focus of research over the past years~\cite{Ref_BH_bassoy2017coordinated}. Second, the synchronization among cooperating base stations have to be accomplished without inter-carrier and inter-symbol interference~\cite{Ref_BH_kotzsch2010interference}. The channel estimation and equalization also must be carried out in a inter-BS-coherent manner, which greatly increases the computation complexity.

\subsubsection{New Modulation}

The 3GPP LTE-Advanced (LTE-A) networks are implemented based on orthogonal frequency division multiple access (OFDMA) \cite{Ref_WJ_jiang2015ofdm}, which is a typical instance of orthogonal multiple access (OMA) technologies prohibiting physical resource block (PRB) sharing by multiple users. In comparison to CDMA, which is deployed in 3G systems, OFDMA shows a conspicuous superiority in combating multi-path fading by simple and robust carrier-based channel equalization. Furthermore, when combined with MIMO, OFDMA is capable to overwhelmingly outperform CDMA in spectral efficiency. Nevertheless, the full performance of MIMO-OFDM highly relies on the MIMO precoding and resource mapping, which have to be precisely adapted to the channel condition to achieve the optimum. As the dimension of MIMO increase, from up to $8\times4$ in LTE-A gradually to over $256\times32$ massive MIMO, and eventually to the future ultra-massive MIMO (e.g. $1024\times64$), the complexity of MIMO-OFDM adaptation is dramatically increasing. Meanwhile, in response to the demand of supporting higher mobility -- which implies higher fading dynamics -- the computation latency constraint to this online adaptation procedure is also becoming more and more strict. To cope with these emerging challenges, a new architecture of AI-driven MIMO-OFDM transceivers has been proposed towards future 6G systems, which relies on AI techniques to efficiently solve the problem of online MIMO precoding and resource mapping~\cite{Ref_BH_zhang20196g}.

Alongside with a further evolution in OFDMA technologies, non-orthogonal multiple access (NOMA) technologies are also widely considered as an answer to the new challenges in the next generation of mobile communication networks. In contrast to OMA, NOMA allow multiple users to reuse the same PRB, which can be achieved by complex inter-user interference cancellation. NOMA approaches can be generally divided into two categories, namely the power-domain (PD) NOMA and the code-domain NOMA. While PD-NOMA has been recently proposed and is attracting a lot of research interest in context of 5G~\cite{Ref_BH_islam2017power}, code-domain NOMA has a longer history in legacy systems (e.g. CDMA in 3G), and provides an alternative to PD-NOMA with numerous variations, such as trellis-coded multiple access (TCMA)~\cite{Ref_BH_aulin1999trellis}, interleave division multiple access (IDMA)~\cite{Ref_BH_li2006interleave}, multi-user shared access (MUSA)~\cite{Ref_BH_yuan2016multi}, pattern-division multiple access (PDMA)~\cite{Ref_BH_chen2017pattern}, and sparse-code multiple access (SCMA)~\cite{Ref_BH_nikopour2013sparse}.

Since the beyond 5G and the 6G networks are expected to simultaneously manage a massive amount of links, e.g. in the mMTC scenario and its future extensions, NOMA solutions appear promising since they provide higher bandwidth efficiency than OMA approaches. Recent studies have also demonstrated that NOMA can be effectively exploited in new spectrum, including mmWave, \si{\tera\hertz}, and optical frequencies. Additionally, when deployed together with CoMP, NOMA has been proven as capable to outperform CoMP-OMA in both power efficiency and spectral efficiency.

Being completely based on successive interference cancellation, NOMA has a significantly higher complexity in its receiver design than OMA, which increases in polynomial or even exponential order along with the number of users. Especially, in some scenarios that require cooperative decoding across different UEs, specific D2D interfaces must be reserved for this functionality, and security/trust concerns shall be taken into account, to enable the deployment of NOMA in 6G.

\subsection{New Architecture}\label{subsec:new_arch}
So far, all legacy and existing cellular systems have been designed to substantially rely on terrestrial base stations. For marine, oceanic, as well as wild terrestrial areas, which are impossible or economically challenging to be covered by terrestrial cellular networks, satellites have been since long the most common communication solution. Aiming at a better coverage rate, deployment of non-terrestrial infrastructures as part of the 6G network is being treated as an emerging topic, known as the integrated space and terrestrial network (ISTN). An ISTN is expected to consist of three layers: the ground-based layer constructed by terrestrial base stations, the airborne layer empowered by HAP and UAV, and the spaceborne layer implemented by satellites. An envision of the architecture for the 6G systems, demonstrating representative use cases and key technological enablers, is illustrated in \figurename \ref{Figure_6Garchitecture}.

\subsubsection{Large-Scale Satellite Constellation}
Until now, the coverage of terrestrial networks has only reached a small portion of the whole surface of the globe. First, it is technically impossible to install terrestrial base stations for offering large-scale coverage in ocean and desert \cite{Ref_WJ_qu2017leo}. Second, it is difficult to cover extreme topographies, e.g., high mountain area, valley, and cliff, while it is not cost-efficient to use a terrestrial network to provide services for sparsely-populated areas. Additionally, terrestrial networks are vulnerable to natural disasters such as earthquake, flood, hurricane, and tsunami, where there is a vital demand of communications but the infrastructure is destroyed or in outage of service. With the expansion of human activity, e.g., the passengers in commercial planes and cruise ships, the demand of MBB services in uncovered areas increasingly grows. Also, the connectivity demand of IoT deployment scenarios like wild environmental monitoring, Offshore wind farm,  and smart grid require wide-area ubiquitous coverage. Satellite communications have been since long the most common solution for wide coverage but currently the mobile  communication service offered by GEO satellites is costly, low data rate, and high latency due to the expensive cost for launching and its wide area coverage (1/3 surface of Earth per GEO satellite).    

The satellites in LEO \cite{Ref_WJ_hu2001satellite} have some advantages over GEO satellite for providing communication services. A LEO satellite operates in an orbit generally lower than $1000\mathrm{km}$, which can substantially lower the latency due to the signal propagation compared to the GEO satellite in the orbit of $36000\mathrm{km}$. Meanwhile, the propagation loss of LEO is much smaller, facilitating the direction connectivity to mobile and IoT devices that are strictly constraint by the battery supply. Moreover, a stationary ground terminal like an IoT device mounted in a monitoring position may suffer from an obstacle in the line of sight from GEO. There is an early attempt to implement a global satellite mobile communication system, i.e., the Iridium constellation \cite{Ref_WJ_grubb1991traveller} that became commercially available in November 1998. It consists of 66 LEO satellites at an altitude of approximately 781 kilometers and provides mobile phone and data services over the entire Earth surface. Even though it fails due to expensive costs and lack of demands at that time, it is a great technological breakthrough. It still operates today and the second generation Iridium system has been successfully deployed last year. 

In recent years, the high-tech company SpaceX gains a lot of attention due to its revolutionary development of space launching technologies.  Its reusable rocket namely Falcon 9 dramatically lower the cost of space launching, opening the possibility for deploying large space infrastructures. In January 2015, SpaceX announced its ambition called Starlink \cite{Ref_WJ_foust2019spacex}, utilizing a very large-scale  constellation with thousands of LEO satellites to provide global Internet access services \cite{Ref_WJ_starlink}. The U.S. FCC has approved its first-stage plan to launch 12~000 satellites and another application for deploying 30~000 additional satellites is under consideration. With the advancement of electronic technology, the weight of each satellite reduces to approximately \SI{260}{\kilo\gram} and a compact, flat-panel design minimizes the volume, allowing for a dense launch of 60 Starlink satellites to make full use of the capability of Falcon 9. Each satellite becomes cheap due to the massive production, in combination with reusable rockets, building  a large-scale constellation becomes feasible from both commercial and technological perspectives.  Since its first launch in May 2019, more than 1000 satellites have been successfully deployed through 17 times launches and the speed of deployment is planned to reach 120 satellites per month.  In October 2020, SpaceX started to invite some early users to join public testing and it was reported that data rate vary from \SI{50}{\mega\bit/\second} to \SI{150}{\mega\bit/\second} and latency from \SI{20}{\milli\second} to \SI{40}{\milli\second} can be expected. Considering its announced monthly fee of 99 USD for active services and its steadily improvement of capacity and performance with the increasing number of deployed satellites, its impact on terrestrial networks should be seriously taken into account. Looking forward to a future global ubiquitous converge that is available anywhere and anytime, it is strongly suggested to integrate satellite networks into the 6G network as part of it.

\subsubsection{High Altitude Platform }
In general, terrestrial networks and satellite communications are two technologies that dominate mobile communications for long years. HAP, a quasi-stationary aerial platform operating in the stratosphere at an altitude between \SIrange{17}{22}{\kilo\meter} above the Earth's surface represents a new alternative to provide a multitude of telecommunication services in a cost-efficient way \cite{Ref_WJ_karapantazis2005role}. 
In comparison with terrestrial base stations, HAP can cover a larger area, offer unobstructed connectivity with high signal arrival angle, and provide the flexibility of quick deployment with less temporal and spatial constraints \cite{Ref_WJ_jiang2014achieving}. Compared to satellite systems, HAP has the following advantages: much lower cost of implementation and deployment due to the avoidance of space launching, the possibility of upgrading, repairing, and redeployment, and much shorter propagation distance that corresponds to higher signal strength and lower latency \cite{Ref_WJ_Mohammed2011role}. 

An aerial platform can carry on-board base stations of both terrestrial and satellite segments, providing connectivity to the terminals of end users transparently. It can keep quasi-stationary, be redeployed, or be moved from one site to another on demand. This is therefore an efficient solution to improve the coverage of terrestrial and satellite systems.  
The aerial platform is not only a host of communication services.  It well adapts to different applications such as high-definition multimedia broadcasting, remote sensing, surveillance, intelligent transportation, and environmental monitoring. It is also applied for navigation and positioning, working standalone or as a local enhancement component of Global Navigation Satellite System, for a higher accuracy and better availability \cite{Ref_WJ_jiang2014opportunistic}. Last but not least, it has  unique value in local temporary events, emergence communications, and natural disaster (e.g., earthquake, flood, and tsunami) relief, where the terrestrial infrastructure is destroyed or the power grid is in outage of services. 

In 6G systems, the synergy among terrestrial, satellite, and HAP is worth exploring to provide ubiquitous, robust, and resilient network infrastructure and communication connectivity. For example, using HAP to provide high-throughput backhaul links for the small cells deployed in the sites that are hard or expensive to provide wired links, or applying HAP to enhance or relay the satellite signals. From the perspective of the integrated terrestrial-HAP-satellite systems, there are several technical challenges to be solved such as the power supply of the aerial platform,  the stability of antenna array, channel models, seamless handover, admission control, interference management, etc. To this end, extensive research on the topics  such as 3D channel modeling \cite{Ref_WJ_dovis2002small}, advanced multi-antenna technologies  \cite{Ref_WJ_Michailidis2010three}, spectrum-awareness, dynamic spectrum management, and FSO \cite{Ref_WJ_fidler2010optical} has been carried.

\subsubsection{Unmanned Aerial Vehicle} Alongside the space satellite and HAP, UAV also play an indispensable role, as the last piece of the puzzle that fills the gap of near-Earth altitude, in the foreseen ISTN.
Over the past years, it has been widely discussed to use UAVs in cellular networks as flying BSs or relays~\cite{Ref_BH_fotouhi2019survey,Ref_BH_wang2019deployment}. Generally, it makes a flexible mobile supplement to the fixed terrestrial gNBs and space satellites, offering a possibility to dynamically re-plan the RAN by flexibly deploying UAVs to different locations. Compared to HAPs, UAVs are deployed closer to the ground, which not only makes them much cheaper, but also grants them better channel gain with lower path loss. Especially,  upon emergencies (e.g. in disaster reliefs or search \& rescue), UAV also provides a low-cost solution of temporary wireless service delivery to inaccessible areas, such like caves, tunnels, and dilapidated buildings in earthquakes/fires, which cannot be covered by satellites and HAPs.

Furthermore, empowered by the latest achievements of wireless power transmission, UAVs can also be exploited as mobile and automated wireless chargers~\cite{Ref_BH_su2020uav}. Especially, with the simultaneous wireless information and power transfer (SWIPT) technologies, the missions of battery charging and information transmission can be accomplished in a seamless joint \cite{Ref_BH_feng2020uav}. This significantly raises the feasibility of massive deployment of battery-life-critical UEs, which is essential in some emerging use scenarios such as dense WSN. With such rich potentials, UAVs are nowadays widely considered as an essential component of future 6G infrastructure.

In addition to its irreplaceable position in the 6G network infrastructure, UAV is also expected to contribute to the prosperity of new use cases and emerging applications in 6G. With its well developed integration with multimedia devices such as video camera and microphone, UAV has since long been widely used in offline photographing and videographing. By nature it is promised to play a larger role in future online video streaming, and can be enabling remote sensing and multi-sense experience when equipped with variant sensors. The high flexibility, mobility and continuously decreasing load cost of UAVs also lead to their foreseeable deployment in future intelligent logistic applications.

\subsection{New Paradigm}
The recent revival of AI technology spurs the discussions of whether 6G will be an integrated system of AI and mobile networks. The 6G system is expected to support the upsurge of diversified mobile AI applications, and in turn AI will play a critical role in designing and optimizing the wireless architecture \cite{Ref_WJ_jiang2020deeplearning}. The similar trends are happening in other fields like blockchain and digital twin, which are recognized as strong drivers to shape the next generation mobile system. It is foreseen that 6G will transform into a huge  computer, which converges distributed communication, computing, storage, sensing, and controlling resources for provisioning services of pervasive computing, AI, blockchain,  digital twin, etc.   

\subsubsection{Artificial Intelligence}
On the list of 6G enabling technologies, AI is recognized as the most potential one. As mobile networks are increasingly complex and heterogeneous, many optimization tasks become intractable, offering an opportunity for advanced ML techniques.  Categorized typically into  supervised, unsupervised, and reinforcement learning, ML is being considered as a promising data-driven tool to provide computational radio and network intelligence from the physical layer \cite{Ref_WJ_jiang2018multi} to network management \cite{Ref_WJ_jiang2017son}. As a sub-branch of ML, deep learning \cite{Ref_WJ_jiang2020deep} can mimic biological nervous systems and automatically extract features, extending across all three mentioned learning paradigms. It has a wide variety of applications to against the big challenges in wireless communications and networking, being applied to form more adaptive transmission (power, precoder, coding rate, and modulation constellation) in massive MIMO \cite{Ref_WJ_huang2019deep}, to enable more accurate estimation and prediction of fading channels \cite{Ref_WJ_jiang2019neural, Ref_WJ_jiang2020recurrent}, to provide a more efficient RF design (pre-distortion for power amplifier compensation, beam-forming, and crest-factor reduction), to deliver a better solution for intelligent network management \cite{Ref_WJ_jiang2017experimental}, and to offer more efficient orchestration for mobile edge computing, networking slicing, and virtual resources management \cite{Ref_WJ_jiang2017intelligent}. 

In addition to deep learning, a few cutting-edge ML techniques represented by federated learning and transfer learning start showing strong potential in wireless communications. Data-driven methods always have to take into account the issue of data privacy, which limits the manner of processing collected data.  In some scenarios, distributing data is strictly prohibited and only local processing on the device where the data was collected is allowed. Federated learning is a method achieving the fulfillment of this requirement by processing the raw data locally and distributing the processed data in a masked form. The mask is designed such that each of the individual data processing expose no information, whereas their cooperation allows for meaningful parameter adjustments towards a universal model.   While federated learning gives a method of training ML models from a large number of data sources without ever exposing sensitive data, it creates only one shared model for universal applicability.  When individual adjustments of models are required for their deployment to be successful, transfer learning can be used as a tool enabling these adjustments and doing so in a manner requiring a much lower amount of data.  By reusing the major part of pre-trained models in a different environment and only adjusting some of the parameters, transfer learning is able to provide quick adaptations using only a low amount of local data. 

In addition to using AI to assist with the operation of the networks (i.e., AI for Networking), it is also important to use the ubiquitous computing, connectivity, storage resources to provide mobile AI services to end users in an AI-as-a-Service paradigm (i.e., Networking for AI) \cite{Ref_WJ_Letaief2019roadmap}. Principally, this provides deep edge resources to enable AI-based computation for new-style terminals such as robots, smart cars, drones, and VR glasses, which demand large amount of computing resources but limited by embedded computing components and power supply.  Such AI tasks mainly means the traditional computation-intensive AI tasks, e.g., computer vision, SLAM, speech and facial recognition, natural language processing, and motion control. 

\subsubsection{Blockchain} 

With the great success of a kind of cryptocurrency known as Bitcoin, the blockchain technology has received enormous attention in both industry and academia \cite{Ref_WJ_tscorsch2016bitcoin}. A blockchain is essentially a distributed public ledger spreading across all participants deployed typically in a peer-to-peer network. A chain of blocks originates from the first block called the genesis block. A new block is appended to the chain via a hash value that is generated according to the information of its parent block.  Each block typically consists of two parts: the block header and transaction data. In particular, the header mainly contains the following information: block version indicating the validation rule, the hash of its parent block, timestamp, the number of transactions, and MerkelRoot that concatenates the hash values of all the transactions in this block. A chain continuously grows as blockchain users perform transactions. A miner records and packs a batch of transactions into a block by solving a computationally difficult problem called Proof of Work (PoW).  The newly mined block is then broadcasted to the whole blockchain network and all the nodes join the consensus process to validate its trustfulness and update the new block into the chain.

The blockchain has the following technological advantages \cite{Ref_WJ_dai2019blockchain}: 
\textit{Immutability}: the transaction data in the blockchain is unchangeable once it is recorded since each block is linked with other blocks via the hash value. The possibility of breaking the whole chain and modifying the content of all blocks is very limited.
\textit{Decentralization}: it applies the consensus mechanisms to manage and maintain the distributed ledge without the need of a centralized entity or third party. The blocks are replicated and shared over an entire blockchain network, thereby avoiding the risk of single point of failure, enhancing data persistency and security, and providing flexibility.  
\textit{Transparency}: all blockchain participants have equal right and can access all transaction information of blockchain.  
\textit{Security} and \textit{privacy}: the adoption of asymmetric cryptography, the inherit feature of data immutability, consensus mechanism, and anonymous addressing ensure the security, trustworthiness, and privacy of the blockchain. Despite these promising merits, scalability is a key barrier when the blockchain technology is widely applied from the perspectives of throughput, storage and networking. Enabling technologies \cite{Ref_WJ_xie2019survey} related to the number of transactions in each block, block interval time, data transmission, and data storage to realize scalable blockchain systems are recently studied.

Recently, the potential of blockchain in 5G and beyond systems has been initially investigated in the literature \cite{Ref_WJ_nguyen2020blockchain}. It is applied to enhance the technologies such as edge computing, NFV, network slicing, and device-to-device communications, to implement important services, e.g., the sharing of spectrum and radio resource, data storage and sharing,  network virtualization, security and privacy, in the use case domains of smart city, smart transportation, smart grid, smart healthcare, and UAV. On the other hand, the deployment of 5G networks can boost the application of blockchain systems. The ubiquitous connectivity, computing, and storage resources provided by mobile networks can be employed to provide local computing power for mobile blockchain systems \cite{Ref_WJ_xiong2018when} so as to support solving PoW puzzles, hashing, encryption, and consensus execution. It is envisioned that the blockchain will be converged into the upcoming 6G system for more flexible, secure, and efficient information infrastructure.

\subsubsection{Digital Twin}
Digital twin is an emerging technology and one significant use case of 6G communication system. It refers to the logical copy (a.k.a., virtual object or softwarized copy) of a physical object \cite{Ref_MAH_Minerva2020DitialTwin}. The virtual representation shall reflect all the dynamics, characteristics, critical components, and important properties of an original physical object that operates and lives throughout its life cycle \cite{Ref_MAH_Pires2019DigitalTwin}. The digital twin is following the life cycle of a physical twin. Therefore, its monitoring, controlling, maintenance, prediction, and optimization processes are started and ended in parallel with its respective physical twin. Each digital twin is linked to its respective physical twin through a unique key. The unique key is used to identify the physical twin and allows to establish a bijective (one-to-one) relationship between the digital twin and its real twin. Recently, digital twin has become the center of attention and has attracted significant attention from the Industry 4.0, research and development community, manufacturing, and others due to its importance in the improvement of the quality of products, services, processes, devices, etc. in a specific context using AI techniques. 

There is a large number of industries such as manufacturing and aviation, which have been developing and commercializing the digital twin in order to optimize their processes since last years \cite{Ref_MAH_Schleich2017DT}. In addition, the digital twin is also in its initial stage in healthcare and medicine fields and is expected to be fully commercialized with the development and deployment of 6G communication networks in the near future \cite{Ref_MAH_Zakrajsek2018DT}. Despite, it is also being extensively applied in the IoT and Industry 4.0 domains. In both scenarios, the AI techniques are used to collect, analyze, and test (in different conditions) the data from a physical object to build its softwarized copy \cite{Ref_MAH_Luscinski2018DT}. The more information about the physical object is provided to the AI analyzer, the more accurate and better the performance and prediction of the virtual object will be during its life cycle \cite{Ref_MAH_Barricelli2019DigitalTwin}. 

Towards full utilization of digital twin in 6G communication system and specifically in the context of Industry 4.0 and IoT, there are still a number of very critical research problems, which are needed to be addressed from both academia and industry. They are, including, but not limited to: dynamically scaling up the platform of digital twin to millions and billions of IoT devices, the deployment of the zero-touch and self-management approaches to the devices and processes, lack of existing models and methodologies in the area of AI to virtualize the physical object, security and privacy, and many others that are thoroughly addressed in \cite{Ref_MAH_Minerva2020DitialTwin}, \cite{Ref_MAH_Pires2019DigitalTwin}, and \cite{Ref_MAH_Barricelli2019DigitalTwin}. These challenges shall be, first and foremost, explored aiming to develop more suitable digital twin solutions for a wide range of IoT deployment scenarios and industrial adaptations in the coming decade. 

\subsubsection{Intelligent Edge Computing} 
Edge computing plays a significant role in increasing the performance of network services, efficient utilization of network resources (both physical and virtual), decreasing CAPEX/OPEX of a mobile operator, and lowering network complexity (both in control plane and user plane) \cite{Ref_MAH_Nasimi2018EdgeIntelligence}. However, the existence of a large number of end users, each with a diverse set of business and technical requirements, challenges the network operator to think on different alternatives in order to address the existing limitations related to edge computing using cutting-edge AI tools and modern ML methods. To that objective, the edge intelligence (EI) is introduced aiming to integrate the AI and ML techniques at the edge of the mobile networks in order to bring automation and intelligence. The EI is envisioned to be one of the key enabling technologies for beyond 5G and 6G communication networks. Thanks to the increasing number of smart portable devices, user equipment, internet of intelligent things (IoIT), and the proliferation of intelligent services; there is a strong demand for the EI in the edge of 6G mobile networks to automate its respective tasks \cite{Ref_MAH_Habibi2020SlicingChapter}. 

One of the major use cases of EI can be the automation of the management and orchestration tasks of the virtual resources in NG-RAN architecture. In this use case, the EI is extended to the NG-RAN in order to automate all tasks related to the RAN network slice subnet management function (NSSMF) and network function management functions (NFMFs) in order to lower the management and orchestration complexity. To that aim, the ETSI has launched the ENI ISG to investigate and provides recommendations to operators \cite{Ref_MAH_ETSI2019EI}. In each of the use cases, including the aforementioned, the EI is constituted of a set of connected devices which are used to collect, normalize, process, and analyze the data. Subsequently, the processed data is sent back to the assisted systems in the form of recommendations and/or orders to be executed in order to automate the target tasks or functionalities \cite{Ref_MAH_ETSI2019EI}. 

The applications of EI in the context of vertical industries have also attracted the attention of both academia and industrial organizations. For example, the EI plays significant role in addressing mission-critical applications and massive and critical mMTC types of services in centralized, semi-centralized, and localized resource allocation scenarios. Despite, the EI also provides energy-efficient solution which are expected to reduce the energy consumption of the communication networks. Therefore, it is considered a novel opportunity for both operators and vertical industries in order to digitize its application and businesses -- using personal computing, fog computing, urban computing, and other mechanisms -- in the 6G mobile networks. 

Despite the aforementioned key advantages, there are still a number of unsolved research challenges in realising EI in beyond 5G mobile networks. It is, therefore, crucial to identify and analyze such open research problems and seek for their theoretical and technical solutions. Among others, the most prominent challenges are: data consistency on every device at edge, data scarcity at edge, bad adaptability of statically trained model, and data privacy and security. We expect that more research efforts are needed to completely realize EI in beyond 5G and 6G networks upon addressing such challenges.

\subsubsection{Communication-Computing-Control Convergence}
Mobile edge networks provide computing and caching capabilities at the network edge, which makes low-latency, high-bandwidth, location-aware pervasive computing services a reality. With the proliferation of IoT and Tactile Internet, a huge number of sensing devices and actuators are connected to mobile networks. The next-generation system is envisioned to become a huge computer that would converge ubiquitous communication, computation, storage, sensing, and controlling as a whole to provide disruptive applications.    

Due to their superiority in integration and mobility over wired connections, wireless links are gradually relied by modern and future controlling systems to close the signal loop, giving birth to the use scenario of URLLC. The spirit of system design behind this concept follows the classic methodology of control \& communication independent design: it starts from designing the controlling component without concerning the characteristics of communication system, where a set of communication requirements will be generated regarding the expected controlling performance; then there follows the designing of wireless system, aiming at achieving the target performance proposed by the last stage. The KPI requirements of URLLC, such as $99.999\%$ reliability and \SI{10}{\milli\second}, were formulated for a generic controlling scenario in this fashion. 
Nevertheless, recent studies have revealed a necessity of in-loop co-designing of communication and controlling systems tightly coupled to each other. For instance, the close-loop reliability of controlling system has been proven exponentially decreasing along with the AoI over the feedback channel~\cite{Ref_BH_ayan2020optimal,Ref_BH_han2020robustness}. Meanwhile, the AoI over control/feedback channels, as a communication metric, is \emph{convex} about the arrival rates of controlling command and feedback information, which correspond to the sampling rate of sensors and decision rate of controller in the control system~\cite{Ref_BH_jiang2020ai}. The performance of communication system, is therefore limited by the design of sensing and controlling systems. 

Similar issues are also to be addressed in cloud computing. While it occasionally happens that some computing task occupies the cloud server for long time, blocking all other pending tasks in the waiting queue and causing a severe congestion, its source may have re-issued the same task with a more up-to-date status, making the previous task outdated and lack of utility. A preemption of the server, i.e. terminating the ongoing task in advance to its completion, will in this case help reduce the age of task and improve the quality of cloud computing service~\cite{Ref_BH_arafa2019timly}. Furthermore, it is also critical for reducing the AoI, to schedule the order of computing tasks from multiple applications to be offloaded to the cloud \cite{Ref_BH_song2019age}. Optimal decisions of such preemption and scheduling, however, cannot be solely solved by the computing server, nor by the communication system, but only achievable in a joint collaboration among the terminal devices, the network controller, and the cloud computing server.

With the dramatically increasing demand for wireless-connected industrial automation, and the irreversible trend that not only most mobile applications but also the mobile networking service itself are gradually cloudified, new approaches are therefore required in 6G to jointly design and optimize this communication-computing-control symbiosis~\cite{Ref_BH_zhao2019toward}.

\section{Conclusions}
This article provided a comprehensive survey on the drivers,  requirements, efforts, and enablers for the next-generation mobile system beyond 5G. It can be concluded that the traditional evolution of a new generation every decade will not terminate at 5G and the first 6G network is expected to be deployed in 2030 or even earlier taking into account great passions of developing 6G from both academia and industry. 6G will accommodate the use cases and applications introduced in 5G such as IoT, Industry 4.0, virtual reality, and automatic driving with better quality of experience in a more cost-efficient, energy-efficient, and resource-efficient manner. Meanwhile, it will enable unprecedented use cases that cannot be supported by 5G, e.g., holographic-type communications, pervasive intelligence, and  global ubiquitous connectability, as well as other disruptive applications that we are unable to yet imagine. 
The trend of mobile communication services expanded from only human centric to connecting also machines and things, started when MTC and IoT were introduced in the age of 5G will continue, and Internet-of-Everything will be realized when 6G comes. The 6G system has to meet extremely stringent requirements on latency, reliability, mobility, and security, as well as provisioning a substantial boost of coverage, peak data rate, user experienced rate, system capacity, and connectivity density, gaining KPIs generally $10$ to $100$ times better in comparison with 5G.

It is envisioned that 6G will take unprecedented transformations that will make it dramatically distinguishing with the previous generations. That is:
\begin{itemize}
\item It will be shifted from a radio communication network based on electronic technologies to a radio-optical system taking advantage of both electronic and photonic technologies so as to exploit the abundant spectral resources in terahertz and visible light bands, especially in indoor optical wireless coverage, to meet the inexorable demand on higher system capacity and peak data rates.
\item It will become a connected intelligent platform that maximizes the synergy between AI and mobile networks. On the one hand, 6G will facilitate the provisioning of over-the-air AI applications, where AI-as-a-Service is offered to end users through pervasive intelligence. On the other hand, AI-driven air interface, algorithms, protocols, and approaches are applied to implement highly-efficient transmission, optimization, control, and management of resources and networks. 
\item With the disruptive advances in large-scale LEO satellite constellation, the 6G system will go beyond terrestrial networks and provide ubiquitous 3D coverage of the whole surface of our planet through an integrated space-aerial-terrestrial network.
\item It will be a smart compute-connect entity by means of converging distributed communication, computing, storage, and big data resources, integrating sensing, localization, and controlling capabilities, and inter-working with emerging paradigms such as AI, blockchain, digital twin, quantum computing and communications.
\item It should be an intelligent, green, sustainable,  and secure system to fully support the informationized and intelligentized society in 2030 and beyond. 
\end{itemize}
 In 1926, engineer and inventor Nikola Tesla stated that ``\textit{When wireless is perfectly applied the whole Earth will be converted into a huge brain}". This prophecy will transform into a reality when 6G comes.  

\bibliographystyle{IEEEtran}
\bibliography{IEEEabrv,Ref_6GReview}
\vfill
\begin{IEEEbiographynophoto}{Wei Jiang} (M'09--SM'19) received the Ph.D. degree in Computer Science from Beijing University of Posts and Telecommunications (BUPT) in 2008. From 2008 to 2012, he was with the 2012 Laboratory, HUAWEI Technologies. From 2012 to 2015, he was with the institute of Digital Signal Processing, University of Duisburg-Essen, Germany. Since 2015, he is a Senior Researcher with German Research Center for Artificial Intelligence (DFKI), which is the biggest European AI research institution and is the birthplace of ``Industry 4.0" strategy. Meanwhile, he is a Senior Lecturer with University of Kaiserslautern, Germany. He is the author of three book chapters and over 60 conference and journal papers, holds around 30 granted patents, and participated in a number of EU and German research projects: \textit{ABSOLUTE}, \textit{5G COHERENT}, \textit{5G SELFNET}, \textit{AI@EDGE}, \textit{TACNET4.0}, and \textit{KICK}. He was the Guest Editor for the Special Issue on ``Computational Radio Intelligence: A Key for 6G Wireless" in \textit{ZTE Communications} (December 2019). He currently serves as an Associate Editor for \textit{IEEE Access} and is a Moderator for \textit{IEEE TechRxiv}.
\end{IEEEbiographynophoto}

\begin{IEEEbiographynophoto}{Bin Han} (M’15) received in 2009 his B.E. degree from Shanghai Jiao Tong University, M.Sc. in 2012 from Technical University of Darmstadt, and in 2016 the Ph.D. degree from Karlsruhe Institute of Technology. Since July 2016 he has been with University of Kaiserslautern as a Senior Lecturer, researching in the broad area of wireless communication and networking, with recent special focus on B5G/6G networks, network slicing, finite blocklength information theory, and information freshness. He is the author of over 30 conference and journal papers, and participated in multiple EU collaborative research projects.
\end{IEEEbiographynophoto} 

\begin{IEEEbiographynophoto}{Mohammad Asif Habibi} received his B.Sc. degree in Telecommunication Engineering from Kabul University, Afghanistan, in 2011. He obtained the M.Sc. degree in Systems Engineering and Informatics from Czech University of Life Sciences, The Czech Republic, in 2016. Since January 2017, he has been working as a Research Fellow and Ph.D. Candidate at Technische Universit\"at Kaiserslautern, Germany. From 2011 to 2014, he joined HUAWEI, where he was working as a Radio Access Network Engineer. His main research interests include Network Slicing, Network Function Virtualization, Resource Allocation, Machine Learning, and Radio Access Network. 
\end{IEEEbiographynophoto} 

\begin{IEEEbiographynophoto}{Hans D. Schotten} (S'93--M'97) received the Ph.D. degrees from the RWTH Aachen University of Technology, Germany, in 1997. From 1999 to 2003, he worked for Ericsson. From 2003 to 2007, he worked for Qualcomm. He became manager of a R\&D group, Research Coordinator for Qualcomm Europe, and Director for Technical Standards. In 2007, he accepted the offer to become the full professor at the University of Kaiserslautern. In 2012, he - in addition - became scientific director of the German Research Center for Artificial Intelligence (DFKI) and head of the department for Intelligent Networks. Professor Schotten served as dean of the department of Electrical Engineering of the University of Kaiserslautern from 2013 until 2017. Since 2018, he is chairman of the German Society for Information Technology and member of the Supervisory Board of the VDE. He is the author of more than 200 papers and participated in 40+ European and national collaborative research projects.
\end{IEEEbiographynophoto}

\end{document}

%% file: MyAcronym2020.tex
\newabbreviation{1g}{1G}{first generation}
\newabbreviation{2d}{2D}{two-dimensional}
\newabbreviation{3d}{3D}{three-dimensional}
\newabbreviation{3g}{3G}{third generation}
\newabbreviation{3gpp}{3GPP}{Third Generation Partnership Project}
\newabbreviation{4g}{4G}{fourth generation}
\newabbreviation{5g}{5G}{fifth generation}
\newabbreviation{5gppp}{5G-PPP}{Fifth Generation Private Public Partnership}
\newabbreviation{6g}{6G}{sixth generation}
\newabbreviation{ai}{AI}{artificial intelligence}
\newabbreviation{aoi}{AoI}{age of information}
\newabbreviation{ar}{AR}{augmented reality}
\newabbreviation{aos}{AoS}{age of synchronization}
\newabbreviation{aes}{AES}{advanced encryption standard}
\newabbreviation{atis}{ATIS}{Alliance for Telecommunications Industry Solutions}
\newabbreviation{capex}{CAPEX}{capital expenditure}
\newabbreviation{cdma}{CDMA}{code division multiple access}
\newabbreviation{comp}{CoMP}{coordinated multi-point}
\newabbreviation{cr}{CR}{cognitive radio}
\newabbreviation{cpu}{CPU}{central processing unit}
\newabbreviation{cu}{CU}{centralized unit}
\newabbreviation{darpa}{DARPA}{Defense Advanced Research Projects Agency}
\newabbreviation{dsa}{DSA}{digital signature algorithm}
\newabbreviation{dlt}{DLT}{distributed ledge technology}
\newabbreviation{dsm}{DSM}{dynamic spectrum management}
\newabbreviation{du}{DU}{distributed unit}
\newabbreviation{e2e}{E2E}{end-to-end}
\newabbreviation{ecc}{ECC}{elliptic curve cryptosystem}
\newabbreviation{ei}{EI}{edge intelligence}
\newabbreviation{embb}{eMBB}{enhanced mobile broad-band}
\newabbreviation{eni}{ENI}{experiential network intelligence}
\newabbreviation{er}{ER}{extended reality}
\newabbreviation{etsi}{ETSI}{European Telecommunication Standard Institute}
\newabbreviation{fcc}{FCC}{Federal Communications Commission}
\newabbreviation{fso}{FSO}{free-space optical}
\newabbreviation{geo}{GEO}{geostationary Earth orbit} 
\newabbreviation{gdpr}{GDPR}{General Data Protection Regulation}
\newabbreviation{gnb}{gNB}{next-generation NodeB}
\newabbreviation{gsm}{GSM}{Global  System  for  Mobile  communications} 
\newabbreviation{hap}{HAP}{high-altitude platform}
\newabbreviation{htc}{HTC}{holographic-type communication}
\newabbreviation{iab}{IAB}{integrated access and backhaul}
\newabbreviation{ict}{ICT}{information and communication technology}
\newabbreviation{idma}{IDMA}{interleave division multiple access}
\newabbreviation{imt}{IMT}{International Mobile Telecommunications}
\newabbreviation{ioe}{IoE}{internet of everything}
\newabbreviation{ioit}{IoIT}{internet of intelligent things}
\newabbreviation{iot}{IoT}{internet of things}
\newabbreviation{ir}{IR}{infrared}
\newabbreviation{irs}{IRS}{intelligent reflecting surfaces}
\newabbreviation{isg}{ISG}{industry specification group}
\newabbreviation{istn}{ISTN}{integrated space and terrestrial network}
\newabbreviation{itur}{ITU-R}{International Telecommunciation Union - Radiocommunication}
\newabbreviation{itut}{ITU-T}{International Telecommunciation Union - Telecommunication}
\newabbreviation{kpi}{KPI}{key performance indicator}
\newabbreviation{lbt}{LBT}{listen-before-talk}
\newabbreviation{ld}{LD}{Laser diode}
\newabbreviation{led}{LED}{light-emitting diodes}
\newabbreviation{leo}{LEO}{low Earth orbit}	
\newabbreviation{los}{LOS}{line of sight}
\newabbreviation{lte}{LTE}{long term evolution}
\newabbreviation{m2m}{M2M}{machine to machine}
\newabbreviation{mbb}{MBB}{mobile broadband}
\newabbreviation{mimo}{MIMO}{multi-input multi-output}
\newabbreviation{ml}{ML}{machine learning}
\newabbreviation{mmtc}{mMTC}{massive machine-type communications}
\newabbreviation{mmwave}{mmWave}{millimeter wave}
\newabbreviation{mulc}{mULC}{massive ultra-reliable low-latency communication}
\newabbreviation{musa}{MUSA}{multi-user shared access}
\newabbreviation{nfmf}{NFMF}{network function management function}
\newabbreviation{nfv}{NFV}{network function virtualization}
\newabbreviation{mano}{MANO}{management and orchestration}
\newabbreviation{ngran}{NG-RAN}{next-generation radio access network}
\newabbreviation{ngmn}{NGMN}{Next   Generation   Mobile   Networks} 
\newabbreviation{nist}{NIST}{National  Institute of  Standards  and  Technology}
\newabbreviation{noma}{NOMA}{non-orthogonal multiple access}
\newabbreviation{nr}{NR}{new radio}
\newabbreviation{nssmf}{NSSMF}{network slice subnet management function}
\newabbreviation{ofdm}{OFDM}{orthogonal frequency-division multipling}
\newabbreviation{oma}{OMA}{orthogonal multiple access}
\newabbreviation{opex}{OPEX}{operational expenditure}
\newabbreviation{oran}{O-RAN}{open radio access network}
\newabbreviation{pd}{PD}{power domain}
\newabbreviation{pdma}{PDMA}{pattern-division multiple access}
\newabbreviation{pnf}{PNF}{physical network function}
\newabbreviation{pop}{PoP}{point of presence}
\newabbreviation{ppp}{PPP}{Public-Private Partnership}
\newabbreviation{prb}{PRB}{physical resource block}
\newabbreviation{qos}{QoS}{quality of service}
\newabbreviation{ran}{RAN}{radio access network}
\newabbreviation{rat}{RAT}{radio access technology}
\newabbreviation{rf}{RF}{radio frequency}
\newabbreviation{ris}{RIS}{reconfigurable intelligent surfaces}
\newabbreviation{rsa}{RSA}{Rivest-Shamir-Adleman}
\newabbreviation{ru}{RU}{radio unit}
\newabbreviation{scma}{SCMA}{sparse-code multiple access}
\newabbreviation{sdn}{SDN}{software-defined networking}
\newabbreviation{slam}{SLAM}{simultaneous localization and mapping}
\newabbreviation{slm}{SLM}{spatial light modulator}
\newabbreviation{sre}{SRE}{smart radio environment}
\newabbreviation{swipt}{SWIPT}{simultaneous wireless information and power transfer}
\newabbreviation{tcma}{TCMA}{trellis-coded multiple access}
\newabbreviation{tdscdma}{TD-SCDMA}{time division synchronous code division multiple access}
\newabbreviation{thz}{THz}{terahertz}
\newabbreviation{uav}{UAV}{unmanned aerial vehicle}
\newabbreviation{ulbc}{ULBC}{ultra-reliable low-latency broadband communication}
\newabbreviation{umbb}{uMBB}{ubiquitous mobile broadband}
\newabbreviation{urllc}{URLLC}{ultra-reliable low-latency communications}
\newabbreviation{vlc}{VLC}{visible light communication}
\newabbreviation{vnf}{VNF}{virtual network function}
\newabbreviation{vr}{VR}{virtual reality}
\newabbreviation{wcdma}{WCDMA}{wideband code division multiple access}
\newabbreviation{wsn}{WSN}{wireless sensor network}
\newabbreviation{wrc}{WRC}{World Radiocommunication Conference}